\documentclass[aps,prd,preprint,a4paper,showpacs,nofootinbib,superscriptaddress]{revtex4-2}
\usepackage{bm}
\usepackage{indentfirst}
\usepackage{amsmath}
\usepackage{graphicx}
\usepackage{float}
\usepackage{amssymb}
\usepackage{subfigure}
\usepackage{amssymb}
\usepackage{hyperref}
\usepackage{epstopdf}
\usepackage[section]{placeins}

\usepackage[utf8]{inputenc}
\hypersetup{
    colorlinks=true,
    linkcolor=red,
    citecolor=blue,
}
\usepackage{color}
\usepackage[T1]{fontenc}
\usepackage{txfonts}
\usepackage{orcidlink}

\usepackage{adjustbox,lipsum}

\newcommand{\Rmnum}[1]{\expandafter\@slowromancap\romannumeral #1@} 
\newcommand{\bq}{\begin{equation}}
\newcommand{\eq}{\end{equation}}
\newcommand{\bqn}{\begin{eqnarray}}
\newcommand{\eqn}{\end{eqnarray}}
\newcommand{\nb}{\nonumber}
\newcommand{\lb}{\label}

\makeatother

\begin{document}
\title{Reflectionless and echo modes in asymmetric Damour-Solodukhin wormholes}

%\author{the authors}

\author{Wei-Liang Qian}\email[E-mail: ]{wlqian@usp.br (corresponding author)}
\affiliation{Escola de Engenharia de Lorena, Universidade de S\~ao Paulo, 12602-810, Lorena, SP, Brazil}
\affiliation{Center for Gravitation and Cosmology, College of Physical Science and Technology, Yangzhou University, Yangzhou 225009, China}
\affiliation{Faculdade de Engenharia de Guaratinguet\'a, Universidade Estadual Paulista, 12516-410, Guaratinguet\'a, SP, Brazil}

\author{Qiyuan Pan}
\affiliation{Key Laboratory of Low Dimensional Quantum Structures and Quantum Control of Ministry of Education, Synergetic Innovation Center for Quantum Effects and Applications, and Department of Physics, Hunan Normal University, Changsha, Hunan 410081, China}
\affiliation{Center for Gravitation and Cosmology, College of Physical Science and Technology, Yangzhou University, Yangzhou 225009, China}

\author{Ramin G. Daghigh}
\affiliation{Natural Sciences Department, Metropolitan State University, Saint Paul, Minnesota, 55106, USA}

\author{Bean Wang}
\affiliation{Department of Physical Sciences and Applied Mathematics, Vanguard University, Costa Mesa, CA 92626, USA}

\author{Rui-Hong Yue}\email[E-mail: ]{rhyue@yzu.edu.cn (corresponding author)}
\affiliation{Center for Gravitation and Cosmology, College of Physical Science and Technology, Yangzhou University, Yangzhou 225009, China}

\begin{abstract}
It is understood that the echo waveforms in ultracompact objects can be regarded as composed mainly of the asymptotic high-overtone quasinormal modes, dubbed echo modes, which predominantly lie parallel to the real frequency axis.  
Alternatively, Rosato {\it et al.} recently suggested that high-frequency quasi-reflectionless scattering modes are primarily responsible for the echo phenomenon.  
This identification relies on greybody factors as stable observables, despite the apparent spectral instability of quasinormal modes.  
In this work, by extending the definition of quasi-reflectionless modes to reflectionless ones and generalizing symmetric Damour-Solodukhin wormholes to asymmetric cases, we examine the underlying similarity between the reflectionless and echo mode spectra in the complex frequency plane.  
Through a primarily analytical treatment, we demonstrate that the asymptotic properties of these two spectra exhibit a strong resemblance, featuring an approximately uniform distribution parallel to the real frequency axis with the same spacing between successive modes. 
Specifically, the real parts of echo modes coincide with those of reflectionless modes at the limit $|\mathrm{Re}\omega| \gg |\mathrm{Im}\omega|$.
While echo modes typically possess non-vanishing imaginary parts, the reflectionless modes of symmetric Damour-Solodukhin wormholes lie precisely on the real frequency axis, with any deviation serving as a measure of the degree of asymmetry of the wormhole. 
We support our derivations by employing two complementary approaches, based on the scattering matrix and the Green's function.
For a given identical source, the waveforms are calculated numerically using the Green's functions.  
The amplitudes of the waveforms associated with reflectionless modes are found to be more pronounced than those of the echo modes, because reflectionless modes typically lie closer to the real frequency axis than the latter.  
It is argued that both perspectives provide effective tools for describing the echo phenomenon.  
\end{abstract}
%\pacs{ 03.75.Dg, 06.30.Gv, 37.25. + k, 91.10.Pp}

%\date{\today}
\date{Dec. 31th, 2025}

\maketitle

\newpage
\section{Introduction}\label{sec1}

Pioneered by Nollert and Price~\cite{agr-qnm-instability-02, agr-qnm-instability-03} and Aguirregabiria and Vishveshwara~\cite{agr-qnm-27, agr-qnm-30}, the black hole spectral instability has aroused much attention in the recent literature.
The essence of such instability resides in the observation that insignificant modifications to the black hole metric, such as an approximation of the Regge-Wheeler effective potential in terms of piecewise functions~\cite{agr-qnm-instability-02, agr-qnm-instability-11} or the introduction of some rather moderate discontinuities, will drastically deform the quasinormal mode~\cite{agr-qnm-review-01, agr-qnm-review-02, agr-qnm-review-03} (QNM) spectrum, particularly the asymptotic behavior of high overtones~\cite{agr-qnm-Poschl-Teller-16}.
This finding undermines the intuitive assertion that an insignificant modification of the effective potential shall not introduce a sizable impact on the resulting QNMs and has been attributed to the non-Hermitian nature~\cite{spectral-instability-review-03, spectral-instability-review-05} of the gravitational system~\cite{agr-qnm-instability-07}.

Specifically, the high overtones of black hole QNMs were found to ascend the imaginary frequency axis~\cite{agr-qnm-continued-fraction-12, agr-qnm-continued-fraction-23}. 
The unique characteristics of QNMs have been understood to be governed solely by the spacetime properties surrounding the black hole, providing crucial information on the spacetime geometry near the event horizon.
Conversely, it has been shown that ``ultraviolet'' metric perturbations~\cite{agr-qnm-instability-07}, characterized by their small spatial scales, can sizably deform the QNM spectrum. 
However, rather than the features near the horizon~\cite{agr-qnm-echoes-01}, the underlying metric perturbations can be planted further away from the compact object~\cite{agr-qnm-echoes-20}, arguably, not being physically insignificant~\cite{agr-qnm-instability-47}.
The emergence of black hole spectral instability, particularly its persistence regardless of the discontinuity's distance from the horizon or its magnitude, somewhat undermines the understanding that the characteristic of QNMs is primarily determined by the spacetime properties surrounding the black hole, whose precise measurements lead to unambiguous information about the spacetime geometry near the event horizon.

Further expanding on these results, Jaramillo {\it et al.}~\cite{agr-qnm-instability-07, agr-qnm-instability-13} systematically explored the implications of spectral stability by analyzing the effects of randomized and sinusoidal perturbations to the metric in terms of the notion of pseudospectrum in the context of black hole perturbation theory.
Their analyses revealed that the boundary of the pseudospectrum moves closer to the real frequency axis, thereby forming a picture of universal instability in high-overtone modes triggered by ultraviolet perturbations.

The significance of spectral instability resides in its observational implications, especially regarding black hole spectroscopy~\cite{agr-BH-spectroscopy-review-04}.
The relevance of the topic is closely connected with the detections of gravitational waves emanating from the binary mergers by the LIGO and Virgo collaboration~\cite{agr-LIGO-01, agr-LIGO-02, agr-LIGO-03, agr-LIGO-04}.
The success of the ground-based facilities has further promoted the ongoing spaceborne projects, such as LISA~\cite{agr-LISA-01}, TianQin~\cite{agr-TianQin-01, agr-TianQin-Taiji-review-01}, and Taiji~\cite{agr-Taiji-01}, leading to speculation that direct observation of ringdown waveforms might also be plausible~\cite{agr-TianQin-05}.
Under realistic astrophysical conditions, the source of gravitational waves is not a single isolated celestial object.
Typically, they are submerged and interacting with surrounding matter. 
This leads to deviations from the ideal metric with perfect symmetries, motivating the studies of the so-called ``dirty'' black holes~\cite{agr-bh-thermodynamics-12, agr-qnm-33, agr-qnm-34, agr-BH-spectroscopy-10} in the context of black hole perturbation theory.
The obtained results on spectral stability warrant further scrutiny of these dirty compact objects.
Regarding the deformed spectrum in the frequency domain, the asymptotic QNMs that align almost parallel to the real axis are closely related to the intriguing concept of echoes, a late-time phenomenon first speculated by Cardoso {\it et al.}~\cite{agr-qnm-echoes-01, agr-qnm-echoes-review-01}. 
As potential observables, echoes were proposed to help distinguish different but otherwise similar gravitational systems via their distinct properties near the horizon. 
Such a physical picture has motivated many studies into echoes across various systems, including exotic compact objects such as gravastars~\cite{agr-eco-gravastar-02, agr-eco-gravastar-03}, boson stars~\cite{agr-eco-gravastar-07}, and particularly wormholes~\cite{agr-wormhole-01, agr-wormhole-02, agr-wormhole-10, agr-wormhole-11, agr-wormhole-43, agr-wormhole-44, agr-wormhole-50, agr-wormhole-51, agr-wormhole-52, agr-wormhole-53, agr-wormhole-54, agr-wormhole-55, agr-wormhole-review-13}. 
Similar to the late-time tail, echoes are also attributed to the analytic properties of Green's function, as analyzed by Mark {\it et al.}~\cite{agr-qnm-echoes-15}. 
In studies of Damour-Solodukhin type wormholes~\cite{agr-wormhole-12}, Bueno {\it et al.}~\cite{agr-qnm-echoes-16} have investigated echoes by explicitly solving for specific frequencies at which the transfer matrix becomes singular, providing further insights into the complex interplay of spacetime geometry and QNMs.
It has been argued~\cite{agr-qnm-instability-65} that the spectral instability is closely associated with the black hole echoes, based on the frequency-domain analyses~\cite{agr-qnm-echoes-20, agr-strong-lensing-correlator-15, agr-qnm-echoes-45}.

The related topic of the spectral instability, echoes, and causality has incited many studies~\cite{agr-qnm-instability-08, agr-qnm-instability-13, agr-qnm-instability-14, agr-qnm-instability-15, agr-qnm-instability-16, agr-qnm-instability-18, agr-qnm-instability-19, agr-qnm-instability-26, agr-qnm-echoes-22, agr-qnm-echoes-29, agr-qnm-echoes-30, agr-qnm-instability-23, agr-qnm-instability-29, agr-qnm-instability-32, agr-qnm-instability-33, agr-qnm-instability-43, agr-qnm-echoes-35,agr-qnm-instability-70, agr-qnm-instability-71}.
Notably, Cheung {\it et al.}~\cite{agr-qnm-instability-15} pointed out that even the fundamental mode can be destabilized under rather generic perturbations.
By introducing a small perturbation to the Regge-Wheeler effective potential, it was shown~\cite{agr-qnm-instability-15} that the fundamental mode undergoes an outward spiral while the deviation's magnitude increases.
In Refs.~\cite{agr-qnm-instability-56, agr-qnm-instability-58}, the numerical findings on the instability of the fundamental mode were interpreted in terms of the instability of an approximated case where the metric perturbation is modeled by an insignificant disjoint potential barrier.
However, if one does not assume that the perturbation is disjoint, some subtlety emerges.
In particular, it was demonstrated~\cite{agr-qnm-instability-55} that, contrary to the existing results, the fundamental mode might be stable, whose validity is confirmed by numerical calculations without introducing any approximation.
This is in agreement with previous findings~\cite{agr-qnm-instability-55}, reinforcing that the low-lying modes are indeed more resilient than the high overtones.

Reflecting on the spectral instability and related to earlier studies~\cite{agr-qnm-instability-18, agr-qnm-67}, it was independently proposed by Oshita~{\it et al.}~\cite{agr-qnm-68, agr-qnm-69, agr-qnm-instability-61} and Rosato~{\it et al.}~\cite{agr-qnm-instability-60} that black hole greybody factors are more robust observables.
Unlike QNMs, it was pointed out that greybody factors remain largely stable against small perturbations to the metric until relatively high frequencies.
Specifically, the greybody factors of a perturbed black hole metric do not significantly deviate from the original black hole.
Nonetheless, deviations from their unperturbed counterparts are only observed at higher (real-valued) frequencies, which can be readily understood by the asymptotic values obtained using the WKB approximation~\cite{agr-qnm-instability-61}.
In~\cite{agr-qnm-instability-60}, the observed stability in the greybody factors was understood as a result of collective interference effects among the aggregate contributions of unstable QNMs. 
Therefore, it was suggested that the greybody factors are more relevant for interpreting ringdown gravitational wave signals.
Mathematically, the greybody can be viewed as constituted by the contribution from the Regge poles~\cite{agr-qnm-Regge-02}.
In terms of the latter, the concept was recently investigated~\cite{agr-qnm-instability-15} along the lines of earlier studies~\cite{agr-qnm-Regge-10, agr-qnm-Regge-11, agr-qnm-Regge-12, agr-qnm-Regge-13}.
On the one hand, the convergent physical quantities obtained at high frequencies~\cite{agr-qnm-instability-15} can be primarily attributed to the validity of the eikonal limit, where the Regge poles become hardly relevant and the low-lying Regge poles are manifestly stable~\cite{agr-qnm-Regge-13}.
On the other hand, at low frequencies, where the contributions from the Regge pole are crucial, numerical calculations indicate that the Regge pole spectrum remains stable.
These findings reinforce the interpretation of the greybody factor's stability given in~\cite{agr-qnm-instability-60}.

More recently, the notion of quasi-reflectionless scattering modes (quasi-RSMs) was introduced in Ref.~\cite{agr-qnm-instability-63}.  
This proposal is motivated by the connection between greybody factors and ringdown waveforms.  
For symmetric cavity potentials, which mimic wormhole-like spacetime configurations, perfect RSMs exist at discrete real frequencies, whereas in more general situations, the minima of the reflectivity correspond to quasi-RSMs.  
As an intriguing feature of the greybody factor in the high-frequency regime, these quasi-RSMs were shown to be primarily responsible for the echo phenomenon of the underlying compact object.  
These modes were found to be approximately uniformly distributed along the real frequency axis.  

Intriguingly, the above properties are reminiscent of those of echo modes discussed in~\cite{agr-qnm-echoes-20, agr-qnm-echoes-45, agr-qnm-instability-65}, even though their definitions are rather different.  
The present study is therefore motivated to explore the underlying similarity between quasi-RSMs and echo modes.  
By promoting the definition of quasi-RSMs to fully reflectionless modes\footnote{After completing this study, we became aware of the work by Tuncer {\it et al.}~\cite{agr-qnm-instability-85}, in which the notion of {\it total absorption} or {\it virtual absorption mode} is developed in the context of complex frequencies. From a physical standpoint, the definition proposed in~\cite{agr-qnm-instability-85} is essentially equivalent to the one adopted here, while the present work focuses on exploring these modes by explicitly incorporating the boundary conditions and the Green’s function.}, these frequencies become complex and correspond to the poles of an appropriately modified Green's function.  
In addition, we investigate asymmetric cavity potentials, specifically asymmetric Damour-Solodukhin wormholes, and demonstrate that the resemblance in the asymptotic properties of the reflectionless and echo-mode spectra can be largely obtained analytically.  
We carry out explicit derivations of the spectra of RSMs and echo modes using two different approaches, based on the evaluation of the scattering amplitudes and on the poles of the modified Green's function.  
In general, the RSMs are found to lie closer to the real frequency axis, leading to more pronounced waveforms associated with RSMs than those related to echo modes for a given identical source.  

The remainder of this paper is organized as follows. 
In Sec.~\ref{sec2}, we briefly review Green's function approach to the black hole QNMs and show that it is readily generalized to the case of reflectionless scattering processes.
Subsequently, in Sec.~\ref{sec3}, we derive the RSMs as complex frequencies using the scattering amplitude for asymmetric Damour-Solodukhin wormholes.
The properties and implications of the solutions are discussed and compared against the echo modes.
An alternative derivation for RSMs is presented in Sec.~\ref{sec4}, using the modified Green's function.
The reminiscence in the two different scenarios is demonstrated in a more transparent fashion.
In Sec.~\ref{sec5}, we illustrate our results by elaborating on a few semi-analytically tractable examples.  
In particular, the waveforms associated with RSMs and echo modes are evaluated and analyzed.  
The concluding remarks are given in the last section.
The complementary mathematical derivations and tecnical details will be relegated to Appx.~\ref{appA},~\ref{appB},~\ref{appC},~\ref{appD} and~\ref{appE}.

\section{Reflectionless modes as an analytic continuation of quasi-reflectionless modes}\label{sec2}

In this section, we first briefly revisit the definition for QNMs employed in black hole perturbation theory, which is then generalized to the reflectionless scattering processes.

We consider that the perturbations in a compact object's metric can be simplified to the following radial part of the master equation~\cite{agr-qnm-review-03},
\begin{eqnarray}
\frac{\partial^2}{\partial t^2}\Psi(t, x)+\left(-\frac{\partial^2}{\partial x^2}+V_\mathrm{eff}\right)\Psi(t, x)=0 ,
\label{master_eq_ns}
\end{eqnarray}
where the spatial coordinate $x$ is known as the tortoise coordinate, and the effective potential $V_\mathrm{eff}$ is governed by the given spacetime metric, spin ${\bar{s}}$, and angular momentum $\ell$ of the waveform.
For instance, it can be the Regge-Wheeler potential $V_\mathrm{RW}$ for the Schwarzschild black hole metric
\bqn
V_\mathrm{eff} = V_\mathrm{RW}=F\left[\frac{\ell(\ell+1)}{r^2}+(1-{\bar{s}}^2)\frac{r_h}{r^3}\right],
\lb{Veff_RW}
\eqn
where 
\bqn
F=1-r_h/r 
\lb{f_RW}
\eqn
is the metric function.  The event horizon radius $r_h=2M$, where $M$ is the black hole mass.  The tortoise coordinate is related to the radial coordinate $r$ by the relation $x=\int dr/F(r)$.

The black hole QNMs are determined by solving the eigenvalue problem defined by Eq.~\eqref{master_eq_ns} in the frequency domain:
\begin{equation}
\frac{d^2\Psi(\omega, x)}{dx^2}+[\omega_n^2-V_\mathrm{eff}(r)]\Psi(\omega, x) = 0 , \label{eq2}
\end{equation}
per the following boundary conditions for asymptotically flat spacetimes
\begin{equation}
\Psi \sim
\begin{cases}
   e^{-i\omega_{n} x}, &  x \to -\infty, \\
   e^{+i\omega_{n} x}, &  x \to +\infty,
\end{cases}
\label{master_bc0}
\end{equation}
which indicates an ingoing wave at the horizon and an outgoing wave at infinity.

The QNMs are governed by the eigenvalues $\omega_{n}$, known as the quasinormal frequencies, where the subscript $n$ represents the overtone number.
They are typically complex numbers attributed to the dissipative nature of Eq.~\eqref{master_bc0}.
These complex frequencies are associated with the analytic properties of the underlying Green's function that satisfies
\begin{equation}
\left[\frac{d^2}{dx^2}+(\omega_{n}^2-V_\mathrm{eff}(r))\right]{G}(\omega, x,y)= \delta(x-y) .\label{DefGreen}
\end{equation}
According to the standard procedure~\cite{agr-qnm-21, agr-qnm-28, agr-qnm-29}, Green's function can be constructed using the form
\begin{equation}
{G}(\omega, x,y)= \frac{1}{W(\omega)}f(\omega, x_<)g(\omega, x_>) ,\label{FormalGreen}
\end{equation}
where $x_<\equiv \min(x, y)$, $x_>\equiv \max(x, y)$, and
\begin{equation}
W(\omega) \equiv W(g, f) = {g} {f}' - {f} {g}' \label{DefWronskian}
\end{equation}
is the Wronskian of $f$ and $g$, 
where $f$ and $g$ are the two linearly independent solutions of the corresponding homogeneous equation satisfying the boundary conditions Eq.~\eqref{master_bc0} at the horizon and infinity.
To be specific, in asymptotically flat spacetimes, $f$ and $g$ possess the following asymptotic forms
\begin{equation}
f(\omega, x) \sim
\begin{cases}
   e^{-i\omega x} &  x \to -\infty \\
   A_{\mathrm{out}}(\omega)e^{+i\omega x}+A_{\mathrm{in}}(\omega)e^{-i\omega x} &  x \to +\infty
\end{cases}
\label{def_f}
\end{equation}
and
\begin{equation}
g(\omega, x) \sim
\begin{cases}
   B_{\mathrm{out}}(\omega)e^{+i\omega x}+B_{\mathrm{in}}(\omega)e^{-i\omega x} &  x \to -\infty\\
   e^{+i\omega x} &  x \to +\infty 
\end{cases}
\label{def_g}
\end{equation}
in asymptotically flat spacetimes, which are bounded at the limit $t\to +\infty$ for $\Im \omega <0$.
In the above expressions, $A_{\mathrm{in}}$, $A_{\mathrm{out}}$, $B_{\mathrm{in}}$, and $B_{\mathrm{out}}$ are the reflection and transmission amplitudes, whose specific forms might be unknown to us but are well-defined for a given metric in principle.
It is noted that the amplitudes of the waveforms satisfy the relations~\cite{book-blackhole-Frolov}
\bqn
B_{\mathrm{out}} &=& A_{\mathrm{in}} ,\nb\\
B_{\mathrm{in}} &=& -A_{\mathrm{out}}^* ,\label{waveformCompleteness}
\eqn
as a result of the waveform's completeness.
Besides, the black hole's reflection and transmission amplitudes are defined as
\bqn
{\mathcal{R}}_\mathrm{BH}(\omega) &=& \frac{B_{\mathrm{in}}}{B_{\mathrm{out}}},\nb\\
{\mathcal{T}}_\mathrm{BH}(\omega) &=& \frac{1}{B_{\mathrm{out}}} ,\label{RefTransA}
\eqn
for an outgoing wave coming from $-\infty$.

The QNMs correspond to the pole of Green's function.
While the intrinsic pole structure primarily comes from the zeros of the Wronskian Eq.~\eqref{DefWronskian}.

Now, one can similarly define the RSM as complex eigenvalues of Eq.~\eqref{eq2} satisfying the boundary condition:\footnote{
The analytic continuation of the frequency to the complex plane should be viewed as a definition of poles of the Green's function Eq.~\eqref{FormalGreenTilde}.
Its usefulness resides in the contour integration, which rewrites the inverse Fourier transform of the time-domain wavefunction (at a given spacetime coordinate) in terms of the sum of residuals, in accordance with Jordan's lemma.
Specifically, for $t>0$, the poles below the real frequency axis are pertinent as they will be included in the contour.}
\begin{equation}
\Psi \sim
\begin{cases}
   e^{+i\omega_{n} x}, &  x \to -\infty, \\
   e^{+i\omega_{n} x}, &  x \to +\infty,
\end{cases}
\label{master_bc3}
\end{equation}
which indicates the incident wave comes from the horizon and traverses the effective potential without being reflected.
The corresponding Green's function has to be modified, but it can be constructed in a similar fashion:
\begin{equation}
\widetilde{G}(\omega, x,y)= \frac{1}{\widetilde{W}(\omega)}\widetilde{f}(\omega, x_<)g(\omega, x_>) ,\label{FormalGreenTilde}
\end{equation}
where
\begin{equation}
\widetilde{f}(\omega, x) \sim
\begin{cases}
   e^{+i\omega x}, &  x \to -\infty, \\
   \widetilde{A}_{\mathrm{out}}(\omega)e^{+i\omega x}+\widetilde{A}_{\mathrm{in}}(\omega)e^{-i\omega x}, &  x \to +\infty ,
\end{cases}
\label{def_tilde_f}
\end{equation}
and
\begin{equation}
\widetilde{W}(\omega) = W(g, \widetilde{f})  . \label{DefWronskianTilde}
\end{equation}

Conversely, when the incident wave comes from spatial infinity and traverses the black hole without suffering any reflection, we have the boundary condition
\begin{equation}
\Psi \sim
\begin{cases}
   e^{-i\omega_{n} x}, &  x \to -\infty, \\
   e^{-i\omega_{n} x}, &  x \to +\infty .
\end{cases}
\label{master_bc4}
\end{equation}
The Green's function can be constructed similarly by defining
\begin{equation}
\widetilde{g}(\omega, x) \sim
\begin{cases}
   \widetilde{B}_{\mathrm{out}}(\omega)e^{+i\omega x}+\widetilde{B}_{\mathrm{in}}(\omega)e^{-i\omega x}, &  x \to -\infty,\\
   e^{-i\omega x}, &  x \to +\infty .
\end{cases}
\label{def_tilde_g}
\end{equation}
Without loss of generality, we will only consider the incident wave coming from the left ($-\infty$) and the solution Eq.~\eqref{def_tilde_f} of the homogeneous equation and the modified Green's function Eq.~\eqref{FormalGreenTilde}.

\section{Reflectionless scattering modes and their asymptotic properties}\label{sec3}

As mentioned, the RSMs can be defined as the analytic continuation of quasi-reflective modes, achieved by promoting the frequency from the real axis to the complex plane.
They correspond to the complex frequencies that asymptotically satisfy the boundary condition Eq.~\eqref{master_bc3}. 
Similar to the QNMs, this definition can be implemented either in terms of the transfer matrix $\mathbb{T}$, when viewed as a scattering problem, or in terms of the poles of the Green's function, when one considers the dynamical evolution of some initial conditions.

In this section, we explore the RSMs as the complex root of the reflectionless condition, namely, specific complex frequencies at which the reflection amplitude vanishes~\cite{agr-qnm-instability-63}.
While following Refs.~\cite{agr-qnm-echoes-16, agr-qnm-instability-63}, we further generalize the discussions to the context of the asymmetric Damour-Solodukhin wormholes.
An essential feature of Damour-Solodukhin wormholes is the echoes, and it has been shown that for symmetric Damour-Solodukhin wormholes, a new branch of QNMs merges and lies parallel to the real frequency axis.
It is understood that they are primarily responsible for the echo waveform~\cite{agr-qnm-echoes-16, agr-qnm-echoes-35}.

In what follows, we derive the RSMs based on the transfer matrix.
While the arguments follow closely those given in~\cite{agr-qnm-echoes-16}, we emphasize that the obtained equation naturally embraces both the QNMs and RSMs of the original black hole and those particular to the wormhole.

\begin{figure}[h]
\begin{minipage}{250pt}
\centerline{\includegraphics[width=1.0\textwidth]{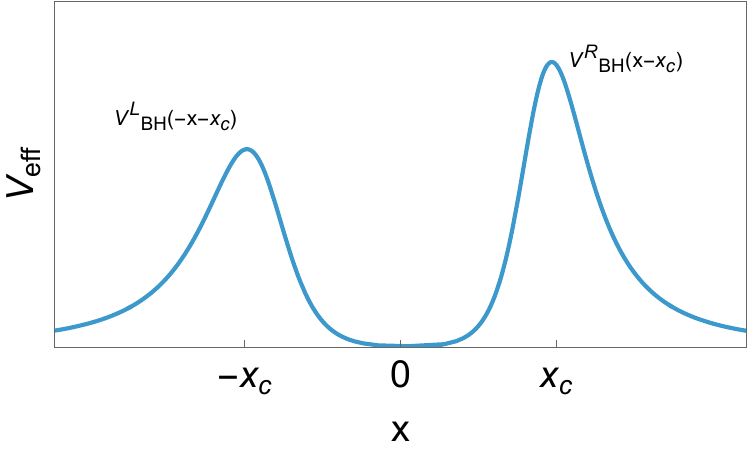}}
\end{minipage}
\renewcommand{\figurename}{Fig.}
\caption{An illustration of the effective potential in an asymmetric Damour-Solodukhin wormhole in the tortoise coordinate $x$.
The effective potential consists of two distinct black hole effective potentials separated by a distance $2x_c$, where the black hole effective potential on the l.h.s. is spatially reflected.}
\label{fig_Veff} 
\end{figure}

As illustrated in Fig.~\ref{fig_Veff}, a Damour-Solodukhin type wormhole can be viewed as constituted by two black hole metrics glued together, whose effective potential has the form
\bqn
V_\mathrm{eff} = V_\mathrm{DS} = V_\mathrm{BH}^\mathrm{L}(-x-x_c) + V_\mathrm{BH}^\mathrm{R}(x-x_c) ,
\eqn
where one of the black hole effective potentials is spatially reflected, and then the two potentials are shifted respectively to the left and right by a distance ($x_c \gg 1$) before being glued at the wormhole's throat ($x=0$). 
When viewed as a scattering problem, the effective potential of each black hole $V_\mathrm{BH}$ gives rise to a $2\times 2$ transfer matrix $T$.
Using the notation and properties given in Eqs.~\eqref{def_f} and~\eqref{waveformCompleteness}, it is found to be
%\bqn
%T =\begin{pmatrix}A_\mathrm{in}^*&A_\mathrm{out}^*\\A_\mathrm{out}&A_\mathrm{in}\end{pmatrix}
%= \begin{pmatrix}B_\mathrm{out}^*&-B_\mathrm{in}\\-B_\mathrm{in}^*&B_\mathrm{out}\end{pmatrix} , \label{TBHmatrix}
%\eqn
\bqn
T 
=\begin{pmatrix}A_\mathrm{in}^*&A_\mathrm{out}\\A_\mathrm{out}^*&A_\mathrm{in}\end{pmatrix}
= \begin{pmatrix}B_\mathrm{out}^*&-B_\mathrm{in}^*\\-B_\mathrm{in}&B_\mathrm{out}\end{pmatrix} , \label{TBHmatrix}
\eqn
which satisfies
\bqn
\begin{pmatrix}C_\mathrm{R}'\\C_\mathrm{L}'\end{pmatrix}
= T \begin{pmatrix}C_\mathrm{R}\\C_\mathrm{L}\end{pmatrix},
\eqn
where the coefficients $C_\mathrm{L, R}$ are the amplitudes of the asymptotic left-going or right-going plane waves:
\begin{equation}
\Psi(x) = 
\begin{cases}
    C_\mathrm{R}'e^{i\omega x} + C_\mathrm{L}'e^{-i\omega x} &\ \ \ \mathrm{for}\ \ \ \  x\to +\infty,\\
    C_\mathrm{R}e^{i\omega x} + C_\mathrm{L}e^{-i\omega x} &\ \ \ \mathrm{for}\ \ \ \  x\to -\infty . 
\end{cases}  
\end{equation}

Given the above construction of a Damour-Solodukhin type wormhole, the corresponding transfer matrix for the entire wormhole metric, for an observer to receive an incident wave from $x\to-\infty$, can be given in terms of individual black holes as a product
\bqn
\mathbb{T}=\mathbb{T}^\mathrm{rhs}\cdot \mathbb{T}^\mathrm{lhs} ,
\eqn
where the matrices are given by
\bqn
\mathbb{T}^\mathrm{rhs}=U^{-1} T^\mathrm{R} U , 
\eqn
and 
\bqn
\mathbb{T}^\mathrm{lhs}= U \sigma \left(T^\mathrm{L}\right)^{-1} \sigma U^{-1} ,\nb
\eqn
where
\bqn
U\equiv U(x_c)=\begin{pmatrix}e^{ +i \omega x_c}&0\\0&e^{-i \omega x_c}\end{pmatrix}
\eqn
is the spatial translation operator and
\bqn
\sigma = \begin{pmatrix}0&1\\1&0\end{pmatrix} .\nb
\eqn
Specifically, the transfer matrix is composed of the black hole effective potential $V_\mathrm{eff}^\mathrm{R}(x)$ and the spatially reflected $V_\mathrm{eff}^\mathrm{L}(-x)$, where the superscript ``R'' and ``L'' denote the corresponding quantities on the r.h.s. and l.h.s. of the wormhole throat.
Moreover, the effective potentials are displaced to the right and left by $x_c$, which is implemented by the phase shifts in the diagonal matrix.
In particular, the transfer matrix for the spatially reflected effective potential $V_\mathrm{eff}^\mathrm{L}(-x)$ is $\sigma T^{-1}\sigma$.
This is because the spatial reflection $x\to -x$ implies not only exchanging the incident and scattered wavefunctions ($T^{-1}$) but also switching the column vector's two components ($\sigma$) twice.
Besides, the effective potential $V_\mathrm{eff}^\mathrm{BH}(-x)$ must be displaced to the left by $x_c$, which is implemented by the phase shift $e^{ i \omega x_c}$ in the diagonal matrix.

Now, let us elaborate on the RSMs.
The corresponding boundary condition Eq.~\eqref{master_bc3} can be achieved by requiring zeros in $\mathbb{T}_{12}$
\bqn
\mathbb{T}_{12}=0 \ \ \ \mathrm{and}\ \ \ \mathbb{T}_{11}\ne 0,\label{T12zero}
\eqn
or poles in $\mathbb{T}_{11}$.\footnote{For complex frequencies, the flux conservation $|\mathcal{R}|^2+|\mathcal{T}|^2=1$ is no longer valid (c.f. Eq.~\eqref{reflecTT} in Appx.~\ref{appA}), and therefore, we do not define reflectionless in terms of the greybody factor $\Gamma\equiv |\mathcal{T}|^2 =1$ as in~\cite{agr-qnm-instability-63}.}

By explicitly writing down the matrix elements, 
\bqn
\mathbb{T}_{12}
&=&e^{2i\omega x_c}T^\mathrm{R}_{11}\left(T^\mathrm{L}\right)_{21}^{-1}+e^{-2i\omega x_c}T^\mathrm{R}_{12}\left(T^\mathrm{L}\right)_{11}^{-1}
=e^{2i\omega x_c}\left[T^\mathrm{R}_{11}\left(T^\mathrm{L}\right)_{21}^{-1}+e^{-4i\omega x_c}T^\mathrm{R}_{12}\left(T^\mathrm{L}\right)_{11}^{-1}\right], \label{T12form}
\eqn
and
\bqn
\mathbb{T}_{11}
&=&T^\mathrm{R}_{11}\left(T^\mathrm{L}\right)_{22}^{-1}+e^{-4i\omega x_c}T^\mathrm{R}_{12}\left(T^\mathrm{L}\right)_{12}^{-1}, \label{T11form}
\eqn
where $\left(T^\mathrm{L}\right)_{ij}^{-1}$ denotes the $ij$th element of the inverse matrix $\left(T^\mathrm{L}\right)^{-1}$.

One immediately concludes that the RSMs are implied by the roots of
\bqn
T^\mathrm{R}_{11}\left(T^\mathrm{L}\right)_{21}^{-1}+e^{-4i\omega x_c}T^\mathrm{R}_{12}\left(T^\mathrm{L}\right)_{11}^{-1} = 0 .\label{case5}
\eqn
This leads to the conclusion that the asymptotic RSMs bear a strong resemblance to those of the echo modes~\cite{agr-qnm-echoes-20}.
To see this, one assumes that the elements of the transfer matrices are moderate functions of the frequency as $\omega\to \infty$.\footnote{These quantities are moderate functions of $\omega$ in the sense that their dependence on the frequency is either not as sensitive as an exponential function in Eq.~\eqref{case0}, or possesses some ondulation of a more extended period. 
In practice, this condition is satisfied given the sufficient condition that the scattering amplitudes (or their ratios) have well-defined values at the limit $\omega\to \infty$.
Also see the examples given in Sec.~\ref{sec5}.}
Therefore, if $\omega$ is a root of Eq.~\eqref{case5} on the complex plane, $\omega+{\pi}/{2x_c}$ will approximately give another root.
These roots asymptotically lie parallel to the real frequency axis, with an interval of $\Delta\omega={\pi}/{2x_c}$.
When inversely transforming back to the time domain, they readily furnish an enveloping factor of the echo waveform of a period $T\sim 2\pi/\Delta\omega=4x_c$.
We note that this feature is also identified later in the Green's function approach by analyzing the poles of Eq.~\eqref{TildeCForm}.

One may refine the above results further by observing that Eq.~\eqref{case5} can be rewritten as
\bqn
T^\mathrm{R}_{11}\left(T^\mathrm{L}\right)_{21}^{-1}\left[1 - e^{-4i\omega x_c}\frac{T^\mathrm{R}_{12}}{T^\mathrm{L}_{12}}\frac{T^\mathrm{L}_{22}}{T^\mathrm{R}_{11}}\frac{\left(T^\mathrm{L}\right)_{12}^{-1}}{\left(T^\mathrm{L}\right)_{21}^{-1}}\right] =  0 , \label{ConsymDS}
\eqn
where one utilizes the definition of the inverse matrix.
For symmetric Damour-Solodukhin wormholes, where the two effective potentials are identical $T^\mathrm{R}=T^\mathrm{L}\equiv T$, this precisely falls back to the case pointed out first in~\cite{agr-qnm-instability-63}: as long as $T^\mathrm{R}_{11}\left(T^\mathrm{L}\right)_{21}^{-1}\ne 0$, a series of purely real roots for $\omega$ is guaranteed.
This is owing to the properties of the transfer matrix Eq.~\eqref{waveformCompleteness} implies that the modulus of all the ratios in the bracket is equal to one.
Subsequently, one only needs to tune the value of $\omega$ on the real frequency axis so that the factor $e^{4i\omega x_c}$ provides an appropriate phase for Eq.~\eqref{ConsymDS} to be valid.
Therefore, there is an infinite number of roots satisfying the properties described above regarding Eq.~\eqref{case5}.
Because they are lying on the real axis, their contribution to the waveform is prominent.
As the symmetry between the two effective potentials disappears, these modes detach from the real frequency axis and collectively migrate into the complex frequency plane, while essentially maintaining their distance from one another.
In this context, the magnitude of the imaginary part of the RSMs measures the deviation from a perfectly symmetric wormhole.
As discussed below, this feature is unique to the RSMs, which is not observed in echo modes.

Besides, it is rather interesting to explore the possible scenarios in which the RSMs are inherited from those of the original black hole.
The first case is that one of the potentials is insignificant when compared to the other.
Without loss of generality, let us assume that $V^\mathrm{R}_\mathrm{eff}$ can be viewed as perturbative.
This indicates that one can approximately take $T^\mathrm{R}_{11} \sim T^\mathrm{R}_{22} \sim 1$ and $T^\mathrm{R}_{12} \sim T^\mathrm{R}_{21} \sim 0$. 
In this case, Eq.~\eqref{T12zero} simplifies to the condition
\bqn
\left(T^\mathrm{L}\right)_{21}^{-1} = 0 ,\label{case3}
\eqn
governing the RSMs of the spatially reflected black hole $V^\mathrm{L}_\mathrm{eff}(-x)$ on the l.h.s. of the throat. 
This is because Eq.~\eqref{case3} implies that a left-going wave incident from the r.h.s. of the potential does not produce any right-going scattered wave on the l.h.s. of the potential.
This is physically intuitive, as the perturbative potential barrier $V^\mathrm{R}_\mathrm{eff}$ will hardly affect a reflectionless incident wave coming from the right.

It is essential to note that, under this scenario, the echo-type modes still merge.
This is because while the magnitude of $T^\mathrm{R}_{12}$ in Eq.~\eqref{case5} is small, it will eventually become appreciable as it will be exponentially amplified by the factor $e^{4i\omega x_c}$ as long as $x_c$ is large enough (given $\mathrm{Im}\omega<0$). 
Following this rationale, the RSMs intrinsic to the wormhole eventually merge when the magnitudes of the two terms in Eq.~\eqref{case5} become comparable.
As $x_c$ further increases, a balance between the two terms is guaranteed by moving the echo-type modes closer to the real axis, while roughly maintaining the product $x_c\mathrm{Im}\omega$ to be a constant, even leading to an interplay between the two types of modes of different origin. 
As discussed below, such a phenomenon has been explored in the context of echo modes.

Another scenario to resort to the original black hole's RSMs corresponds to the solution 
\bqn
T^\mathrm{R}_{12} = \left(T^\mathrm{L}\right)_{21}^{-1} = 0 ,
\eqn
or 
\bqn
T^\mathrm{R}_{11}=\left(T^\mathrm{L}\right)_{11}^{-1} = 0 .
\eqn
The first equality implies the coincidence when the frequencies of the RSMs of the black hole potential on the l.h.s. of the throat are identical to those of the black hole on the r.h.s.
Similarly, the second equality corresponds to the case when the negative values of the quasinormal frequencies of the black hole potential on the r.h.s. of the throat coincide with the QNMs of the black hole on the l.h.s.
In most cases, the above equations do not have solutions.

One is left to explore the case when RSMs are attributed to possible divergence of $\mathbb{T}_{11}$, namely,
\bqn
\frac{\mathbb{T}_{11}}{\mathbb{T}_{12}}=\infty .\label{T11div}
\eqn
Specifically, Eq.~\eqref{T11div} means the RSMs are dictated by the poles in $\mathbb{T}_{11}$ where $\mathbb{T}_{12}$ remains finite or $\mathbb{T}_{11}$ diverges faster than $\mathbb{T}_{22}$ at the quasinormal frequencies.
Observing Eqs.~\eqref{T12form} and~\eqref{T11form}, this happens when either $\left(T^\mathrm{L}\right)_{12}^{-1}$ or $\left(T^\mathrm{L}\right)_{22}^{-1}$ becomes divergent while both $\left(T^\mathrm{L}\right)_{11}^{-1}$ and $\left(T^\mathrm{L}\right)_{21}^{-1}$ remain finite.
However, due to the relation Eq.~\eqref{waveformCompleteness}, the above condition is never satisfied.

For comparison purposes, let us also revisit the evaluation of the QNMs. 
We may enforce the outgoing wave boundary condition Eq.~\eqref{master_bc0} by requiring either the zeros in $\mathbb{T}_{22}$
\bqn
\mathbb{T}_{22}=0 \ \ \ \mathrm{and}\ \ \ \mathbb{T}_{21}\ne 0,\label{T22zero}
\eqn
or the poles in $\mathbb{T}_{21}$.

It is straightforward to show that 
\bqn
\mathbb{T}_{22}
&=&e^{4i\omega x_c}T^\mathrm{R}_{21}\left(T^\mathrm{L}\right)_{21}^{-1}+T^\mathrm{R}_{22}\left(T^\mathrm{L}\right)_{11}^{-1}
%=e^{2i\omega x_c}\left[T^\mathrm{R}_{21}\left(T^\mathrm{L}\right)_{21}^{-1}+e^{-4i\omega x_c}T^\mathrm{R}_{22}\left(T^\mathrm{L}\right)_{11}^{-1}\right] ,
\label{T22form}
\eqn
and
\bqn
\mathbb{T}_{21}
&=&e^{2i\omega x_c}T^\mathrm{R}_{21}\left(T^\mathrm{L}\right)_{12}^{-1}+e^{-2i\omega x_c}T^\mathrm{R}_{22}\left(T^\mathrm{L}\right)_{22}^{-1} .
\label{T21form}
\eqn

Eq.~\eqref{T22form} immediately implies a branch of modes whose asymptotic properties are associated with the roots of
\bqn
e^{4i\omega x_c}T^\mathrm{R}_{21}\left(T^\mathrm{L}\right)_{21}^{-1}+T^\mathrm{R}_{22}\left(T^\mathrm{L}\right)_{11}^{-1} = 0 ,\label{case0}
\eqn
on the complex plane, which asymptotically lie parallel to the real axis~\cite{agr-qnm-echoes-20}, with an interval of $\Delta\omega={\pi}/{2x_c}$.
When inversely transforming back to the time domain, they readily furnish the echo waveform.

Now, it is attempting to derive an expression for QNMs similar to Eq.~\eqref{ConsymDS} in the case of a symmetric Damour-Solodukhin wormhole, where the two effective potentials are identical.
However, this is not possible here.
To be more specific, one instead finds 
\bqn
T^\mathrm{R}_{21}\left(T^\mathrm{L}\right)_{21}^{-1}\left[1 - e^{-4i\omega x_c}\frac{T^\mathrm{R}_{22}T^\mathrm{L}_{11}}{T^\mathrm{R}_{21}T^\mathrm{L}_{21}}\right] =  0 . \label{ConsymDS2}
\eqn
At large frequencies, one typically has $\left| T_{21} / T_{22} \right|\ne 1$.
Unlike Eq.~\eqref{ConsymDS}, Eq.~\eqref{ConsymDS2} actually implies that the real part of the echo modes usually approaches a non-vanishing constant in a symmetric Damour-Solodukhin wormhole.
Also, we note that the fundamental mode as the root of Eq.~\eqref{ConsymDS2} with the smallest imaginary part is not related to the original black hole.

Also, it is interesting to explore the scenarios in which the QNMs of the original black hole persist.
The first case is that one of the potentials is insignificant when compared to the other.
Without loss of generality, we assume that $V^\mathrm{R}_\mathrm{eff}$ can be viewed as perturbative.
This indicates that the reflection amplitude will be insignificant, and from Eq.~\eqref{T22form}, this leads to the condition
\bqn
\left(T^\mathrm{L}\right)_{11}^{-1} = 0 ,\label{case1}
\eqn
which is nothing but the QNMs of the original black hole potential $V^\mathrm{L}_\mathrm{eff}$.
We note that under this scenario, the echo modes actually still merge ``dynamically'' from and interact with the original black hole's QNMs.
This is because while the magnitude of $T^\mathrm{R}_{12}$ in Eq.~\eqref{case0} is small, it will eventually become appreciable as the other term will be exponentially suppressed by the factor $e^{-4i\omega x_c}$ once $x_c$ becomes large enough. 
The echo modes eventually merge when the magnitudes of the two terms in Eq.~\eqref{case0} become comparable.
As $x_c$ further increases, a balance between the two terms in Eq.~\eqref{case0} will be maintained, as the echo modes move closer to the real frequency axis, as elaborated in the literature~\cite{agr-qnm-instability-65, agr-qnm-Poschl-Teller-17}. 
In particular, the echo modes have been observed to interplay with the fundamental mode of the original black hole and, eventually, they will dominate the spectrum and also demonstrate themselves in the time-domain waveform (c.f. Fig.~1 of~\cite{agr-qnm-instability-65}).
In some particular case, the dynamical evolution of the echo modes has also been observed, characterized by a bifurcation point that merges in the QNM spectrum (c.f. Fig.~6 of~\cite{agr-qnm-Poschl-Teller-17}).

Similarly, another scenario unlikely to happen resorts to the original black hole's QNMs that correspond to the solution 
\bqn
T^\mathrm{R}_{21}=\left(T^\mathrm{L}\right)_{11}^{-1} = 0 ,\label{rQNM}
\eqn
or 
\bqn
T^\mathrm{R}_{22} = \left(T^\mathrm{L}\right)_{21}^{-1} = 0 .
\eqn
The first equality implies the coincidence when the frequency of the QNMs of the black hole potential on the l.h.s. of the throat is identical to that of the RSMs of the black hole on the r.h.s.
Similarly, the second equality corresponds to the case when the frequency of the QNMs of the black hole potential on the r.h.s. of the throat is identical to that of the RSMs of the black hole on the l.h.s., but for a left-going incident wave from the right.
These modes, even when they exist, have nothing to do with the echo modes.
Although it is possible if one of the potentials is purely reflectionless~\cite{condensed-RSM-10}, for most cases, the above equations do not possess nontrivial solutions.

The last possibility is when QNMs are attributed to the poles of $\mathbb{T}_{21}$ while $\mathbb{T}_{22}$ remains finite, namely, $\frac{\mathbb{T}_{21}}{\mathbb{T}_{22}}=\infty$.
Observing Eqs.~\eqref{T22form} and~\eqref{T21form}, this might happen when either $\left(T^\mathrm{L}\right)_{12}^{-1}$ or $\left(T^\mathrm{L}\right)_{22}^{-1}$ becomes divergent while both $\left(T^\mathrm{L}\right)_{21}^{-1}$ and $\left(T^\mathrm{L}\right)_{11}^{-1}$ remain finite, which is again not possible by the properties of the transfer matrix.

\section{An altnernative approach based on Green's function}\label{sec4}

As discussed, a general recipe for echoes from exotic compact objects via Green's function was proposed in~\cite{agr-qnm-echoes-15}.
In this section, after giving an account of the scheme, we explicitly evaluate the reflection amplitude in the context of a wormhole metric and assess the resulting QNM and RSM modes. 
In particular, we elaborate on a seeming dilemma between the results from the Green function approach and those obtained in the last section.

We first briefly review Green's function approach for echoes in exotic compact objects in the context of wormhole metric primarily based on~\cite{agr-qnm-echoes-15}.
In a Damour-Solodukhin wormhole, the event horizon of the original black hole metric is replaced by a thin shell.
The effective potential near the throat essentially has the shape of a flat valley and vanishes $V_c \sim 0$, and the throat's coordinate will be chosen as $x=0$.
At the throat, when compared to the black hole metric, the ingoing waveform will pick up a fraction of the outgoing wave.
Specifically, we have
\bqn
f_3(\omega,x) = \mathcal{D}(\omega)\left[ f_1(\omega,x)+{\mathcal{C}}(\omega)g_1(\omega,x) \right] ,\label{h1Cform}
\eqn
where ${\mathcal{D}}(\omega)$ is a normalization factor.
Eq.~\eqref{h1Cform} is essentially a combination of the ingoing and outgoing waveforms, whose asymptotical forms are given by 
\bqn
f_1(\omega, x)=\left\{\begin{matrix}e^{-i \omega(x-x_c)}&x\to \ \mathrm{throat} \\A_{\mathrm{out}}e^{i\omega (x-x_c)}+A_{\mathrm{in}}e^{-i\omega (x-x_c)} &x\to +\infty\end{matrix}\right. ,\label{Asf1}
\eqn
and
\bqn
g_1(\omega, x)=\left\{\begin{matrix}e^{i\omega (x-x_c)}&x\to +\infty \\B_{\mathrm{out}}e^{i \omega (x-x_c)}+B_{\mathrm{in}}e^{-i \omega (x-x_c)} &x\to  \ \mathrm{throat}\end{matrix}\right. ,\label{Asg1}
\eqn
where we have denoted the limit for the ingoing wave at $x\to \ \mathrm{throat}$ instead of $x\to -\infty$.
When compared with Eqs.~\eqref{def_f} and~\eqref{def_g}, they are essentially the spatially translated solutions of the black hole's master equation that are asymptotically plane waves.
As the effective potential of the original black hole is shifted, the waveforms must be measured w.r.t. the coordinate $x_c$.

From the perspective of a reflection process that occurred at $x=x_c$, one can write down the ingoing waveform $f_3$ as a linear combination of two plane waves 
\bqn
f_3(\omega,x)\propto e^{-i\omega (x-x_c)} +{\mathcal{R}}(\omega)e^{i\omega (x-x_c)} ,\label{h1Rform}
\eqn
for $x\to \ \mathrm{throat}$, where the reflection amplitude ${\mathcal{R}}$ is dictated mainly by the specific nature of the compact object.
The coefficient $\mathcal{C}(\omega)$ in Eq.~\eqref{h1Cform} is found to be
\bqn
{\mathcal{C}}(\omega) = \frac{{\mathcal{T}}^\mathrm{R}_\mathrm{BH}(\omega){\mathcal{R}}(\omega)}{1-{\mathcal{R}}^\mathrm{R}_\mathrm{BH}(\omega){\mathcal{R}}(\omega)} ,\label{relCR}
\eqn
where one makes use of the asymptotical forms given by Eqs.~\eqref{Asf1} and~\eqref{Asg1} at $x\to \ \mathrm{throat}$, then compares Eq.~\eqref{h1Cform} against Eq.~\eqref{h1Rform} and equates the ratios of the coefficients before the terms $e^{\pm i\omega x}$, while making use of the definitions Eqs.~\eqref{RefTransA}.

One plugs $f_3$ (Eq.~\eqref{h1Cform}) and $g_1$ (Eq.~\eqref{Asg1}) into the definition of Green's function Eq.~\eqref{FormalGreen}, and for $y<x$ we have~\cite{agr-qnm-echoes-15}
\bqn
{G}(\omega,x,y) 
&=& \frac{f_3(\omega,y) g_1(\omega,x)}{W(g_1,f_3)}\nb\\
%&=& \frac{f_1(\omega,y) g_1(\omega,x)}{W(g_1,f_1)}+{\mathcal{C}}(\omega)\frac{g_1(\omega,y) g_1(\omega,x)}{W(g_1,f_3)}\nb\\
&=& \frac{f_1(\omega,y) g_1(\omega,x)}{W(g_1,f_1)}+{\mathcal{C}}(\omega)\frac{g_1(\omega,y) g_1(\omega,x)}{W(g_1,f_1)}\nb\\
&\equiv& {G}^\mathrm{R}_\mathrm{BH}(\omega,x-x_c,y-x_c) +{\mathcal{C}}(\omega)\frac{g_1(\omega,y) g_1(\omega,x)}{W^\mathrm{R}_\mathrm{BH}(\omega)} ,\label{Gtilde_h3}
\eqn
where the factor ${\mathcal{D}}(\omega)$ apparently cancels out and $W^\mathrm{R}_\mathrm{BH}=W(g, f)=W(g_1,f_1)$ and ${G}^\mathrm{R}_\mathrm{BH}$ are the Wronskian and (the spatially translated) Green's function of the original black hole metric on the r.h.s. of the throat.

From a rather generic perspective, the presence of echoes has been attributed to the phase shift of the reflection amplitude ${\mathcal{R}(\omega)}$~\cite{agr-qnm-echoes-15, agr-strong-lensing-correlator-15}.
We note that particular care must be taken so that the definitions of ${\mathcal{C}}$ and ${\mathcal{R}}$ only reflect the physical nature of the compact object and must be independent of the specific choice of coordinates.

For the case of Damour-Solodukhin wormholes, the reflection amplitude can be evaluated explicitly.
By making use of the transfer matrix and the definition~\eqref{h1Rform}, for QNMs, the wormhole's reflection amplitude is found to be
\bqn
{\mathcal{R}}(\omega)
=\frac{\mathbb{T}^\mathrm{WH}_{21}}{\mathbb{T}^\mathrm{WH}_{22}} 
=\frac{\left(T^\mathrm{L}\right)^{-1}_{12}}{\left(T^\mathrm{L}\right)^{-1}_{11}}e^{4i\omega x_c}
\equiv \bar{\mathcal{R}}^\mathrm{L}_\mathrm{BH}e^{4i\omega x_c}.\label{TildeRForm}
\eqn
where
\bqn
\mathbb{T}^\mathrm{WH}= U\mathbb{T}^\mathrm{lhs}U^{-1} ,\label{def_TWH}
%=U^2 \sigma \left(T^\mathrm{L}\right)^{-1} \sigma U^{-2} ,
\eqn
involves an additional spatial translation.
Unlike Eq.~\eqref{RefTransA}, this reflection amplitude corresponds to the scattering of an ingoing wave expressed w.r.t. to the coordinate $x=x_c$.
As first pointed out in~\cite{agr-qnm-echoes-15}, the factor $e^{-4i\omega x_c}$ in $\mathcal{R}(\omega)$ gives rise to the echo period 
\bqn
T = 4x_c ,\label{echoPeriodWH}
\eqn
which measures the period for a wavepacket to bounce between $-x_c$ and $x_c$.

Specifically, by substituting Eq.~\eqref{TildeRForm} into Eq.~\eqref{relCR}, one finds
\bqn
 {\mathcal C}(\omega)
=\frac{{\mathcal{T}}^\mathrm{R}_\mathrm{BH}\bar{\mathcal{R}}^\mathrm{L}_\mathrm{BH}}{e^{-4i \omega x_c} - {\mathcal{R}}^\mathrm{R}_\mathrm{BH}\bar{\mathcal{R}}^\mathrm{L}_\mathrm{BH}} . \label{TildeCForm}
\eqn
As a further confirmation, an alternative approach to derive Eqs.~\eqref{TildeRForm} and~\eqref{TildeCForm} is given in Appx.~\ref{appB}.
The origin of echoes stems from the inverse Fourier transform of ${\mathcal C}(\omega)$
\bqn
\mathcal{C}(t) = \int_{-\infty}^{+\infty} \frac{d\omega}{2\pi}{\mathcal{C}}(\omega)e^{-i\omega t}
=A\left[\delta(t)+B\delta(t-T)+B^2\delta(t-2T)\cdots\right],\label{InvrelCR}
\eqn
where
\bqn
A &=& -\frac{{\mathcal{T}}^\mathrm{R}_\mathrm{BH}}{{\mathcal{R}}^\mathrm{R}_\mathrm{BH}},\nb\\
B &=& \frac{1}{{\mathcal{R}}^\mathrm{R}_\mathrm{BH}{\bar{\mathcal{R}}}^\mathrm{L}_\mathrm{BH}} .
\eqn
Subsequently, the time domain profile governed by the second term on the r.h.s. of Eq.~\eqref{Gtilde_h3} is a convolution of ${\mathcal{C}}(t)$ and the remainder factor $\mathcal{H}(t)$
\bqn
G(t,y,x) \sim \int d\tau \mathcal{H}(t) \mathcal{C}(t-\tau)
=\mathcal{H}(t)+\mathcal{H}(t-T)+\mathcal{H}(t-2T)+\cdots,
\eqn
featuring echoes of period $T$, where
\bqn
\mathcal{H}(t) = \int_{-\infty}^{+\infty} \frac{d\omega}{2\pi}\frac{g_1(\omega,y) g_1(\omega,x)}{W^\mathrm{R}_\mathrm{BH}}e^{-i\omega t} .
\eqn

Similarly, for RSMs, the wormhole's reflection amplitude is
\bqn
{\mathcal{R}}(\omega)
=\frac{\mathbb{T}^\mathrm{WH}_{11}}{\mathbb{T}^\mathrm{WH}_{12}} 
=\frac{\left(T^\mathrm{L}\right)^{-1}_{22}}{\left(T^\mathrm{L}\right)^{-1}_{21}}e^{4i\omega x_c}
= \frac{1}{\bar{\mathcal{R}}^\mathrm{L*}_\mathrm{BH}}e^{4i\omega x_c} ,\label{TildeForm4RSM}
\eqn
which leads to
\bqn
 {\mathcal C}(\omega)
=\frac{{\mathcal{T}}^\mathrm{R}_\mathrm{BH}/\bar{\mathcal{R}}^\mathrm{L*}_\mathrm{BH}}{e^{-4i \omega x_c} - {\mathcal{R}}^\mathrm{R}_\mathrm{BH}/\bar{\mathcal{R}}^\mathrm{L*}_\mathrm{BH}} . \label{TildeCForm4RSM}
\eqn
The remaining arguments for the presence of echoes are mainly identical to the case of QNMs.

However, a subtle dilemma arises in the above reasoning.
More precisely, both denominators in the last line of Eq.~\eqref{Gtilde_h3} involve the Wronskian of the original black hole.
At face value, this seems to suggest that the QNMs of the original black hole would carry over unchanged into the wormhole geometry.
Such a conclusion, however, is atypical, as the spectrum of the black hole is not usually inherited by the wormhole.
From a practical perspective, neither the frequency-domain calculations (see the numerical results in the following section) nor the time-domain simulations provide any indication of these inherited modes.

At first glimpse, one might argue that the singularity caused by the Wronskian $W^\mathrm{R}_\mathrm{BH}$ might be canceled out identically by the normalization factor $\mathcal{D}(\omega)$ defined in Eq.~\eqref{h1Cform}.
However, since $f_3$ appears both in the numerator and denominator in the definition of the Green's function, this factor always cancels out precisely.
As a result, it seems that the vanishing Wronskian $W^\mathrm{R}_\mathrm{BH}$ will furnish a pole to the Green's function Eq.~\eqref{h1Cform}.
Moreover, as its presence is irrelevant to the boundary condition, it seems that those frequencies should appear for both the QNMs and RSMs of the resulting wormhole.
The above statement is incorrect due to an oversight of the following subtlety associated with the properties of the normalization factor.
The factor $\mathcal{D}(\omega)$ is a function of frequency (but not the spatial coordinate), and it indeed becomes divergent as the Wronskian vanishes. 
Besides, the product $\mathcal{D}(\omega)W^\mathrm{R}_\mathrm{BH}$ has a well-defined limit as the frequency attains the original black hole's QNMs.
However, despite the above divergence of $\mathcal{D}(\omega)$, the numerator of the Green's function, shown on the r.h.s. of the first line of Eq.~\eqref{h1Cform} does not become divergent, even though either of the two terms (namely, $\mathcal{D}(\omega)f_1(\omega, x)$ and $\mathcal{D}(\omega)\mathcal{C}(\omega)g_1(\omega, x)$) that constitute $f_3$ manifestly does.
This is because the divergence in the above two terms cancels out identically, which can be shown by explicit calculations given in Appx~\ref{appC}.
To be more specific, the second term, $\mathcal{D}(\omega)\mathcal{C}(\omega)g_1(\omega, x)$, which does not play a part in the denominator of the Green's function, since $W(g_1, g_1)=0$, is present in the numerator and assists a crucial role to cancel out the divergence caused by the troublesome normalization factor.
The third line of Eq.~\eqref{Gtilde_h3} is indeed correct and can be employed to account for both the echo modes and the RSM through the coefficient $\mathcal{C}(\omega)$.  
However, its formulation may introduce some confusion because of the poles arising from the vanishing Wronskian $W^\mathrm{R}_\mathrm{BH}$ in both terms.  
As clarified in the above discussion, these poles in the Green’s function, associated with the original black hole, are irrelevant, and only the poles stemming from the coefficient $\mathcal{C}(\omega)$ remain physically pertinent.

\section{Numerical examples}\label{sec5}

In this section, we present a few examples to illustrate the results obtained in the previous sections.  
First, in Sec.~\ref{sec5.1}, we evaluate and compare the spectra of quasi-RSMs, RSMs, and echo modes in two toy-model wormholes.  
Subsequently, in Sec.~\ref{sec5.2} we compute and compare the corresponding time-domain waveforms for a given source term, exploring the differences that are closely connected to the underlying pole structures. 

\subsection{The spectra of quasi-RSMs, RSMs, and echo modes}\label{sec5.1}

We consider two toy-model wormholes.
The first example is the double-$\delta$ potential of the form
\bqn
V_\mathrm{BH}^\mathrm{L,R}(x) = V_0^\mathrm{L,R}\delta(x) .
\eqn
For each $\delta$-function potential barrier, it is straightforward to show that the transfer matrices are
\bqn
T 
=\begin{pmatrix}1-i\frac{V_0^\mathrm{L,R}}{2\omega}&-i\frac{V_0^\mathrm{L,R}}{2\omega}\\ i\frac{V_0^\mathrm{L,R}}{2\omega}&1+i\frac{V_0^\mathrm{L,R}}{2\omega}\end{pmatrix} , \label{Tdelta}
\eqn
and
\bqn
T^{-1} 
=\begin{pmatrix}1+i\frac{V_0^\mathrm{L,R}}{2\omega}&i\frac{V_0^\mathrm{L,R}}{2\omega}\\ -i\frac{V_0^\mathrm{L,R}}{2\omega}&1-i\frac{V_0^\mathrm{L,R}}{2\omega}\end{pmatrix} . \label{TdeltaInv}
\eqn

For the RSMs, Eq.~\eqref{ConsymDS} gives
\bqn
T^\mathrm{R}_{11}\left(T^\mathrm{L}\right)_{21}^{-1}\left[1 - e^{-4i\omega x_c}\frac{T^\mathrm{R}_{12}}{T^\mathrm{L}_{12}}\frac{T^\mathrm{L}_{22}}{T^\mathrm{R}_{11}}\frac{\left(T^\mathrm{L}\right)_{12}^{-1}}{\left(T^\mathrm{L}\right)_{21}^{-1}}\right] =
-\left(1-i\frac{V_0^\mathrm{R}}{2\omega}\right)i\frac{V_0^\mathrm{L}}{2\omega}\left[1 + e^{-4i\omega x_c}\frac{V_0^\mathrm{R}}{V_0^\mathrm{L}}\frac{1+i\frac{V_0^\mathrm{L}}{2\omega}}{1-i\frac{V_0^\mathrm{R}}{2\omega}}\right] =  0 . \label{EqRSM_DSdelta}
\eqn
For asymptotical RSMs $\omega_n\to \infty$ in Damour-Solodukhin wormholes, the above equation simplifies to
\bqn
\frac{V_0^\mathrm{L}}{V_0^\mathrm{R}} + e^{-4i\omega x_c}  = 0 ,\label{asyDeltaRSM}
\eqn
with complex roots
\bqn
\omega_n \simeq \left(n+\frac12\right)\frac{\pi}{2x_c} - i \frac{1}{4x_c} \ln\left(\frac{V_0^\mathrm{L}}{V_0^\mathrm{R}}\right) ,\label{rootRSM}
\eqn
which falls back to purely real ones for symmetric Damour-Solodukhin wormholes where $V_0^\mathrm{L}=V_0^\mathrm{R}$
\bqn
\omega_n \simeq \left(n+\frac12\right)\frac{\pi}{2x_c} .\label{rootRSMread}
\eqn

\begin{figure}[ht]
\centerline{
\includegraphics[height=0.35\textwidth]{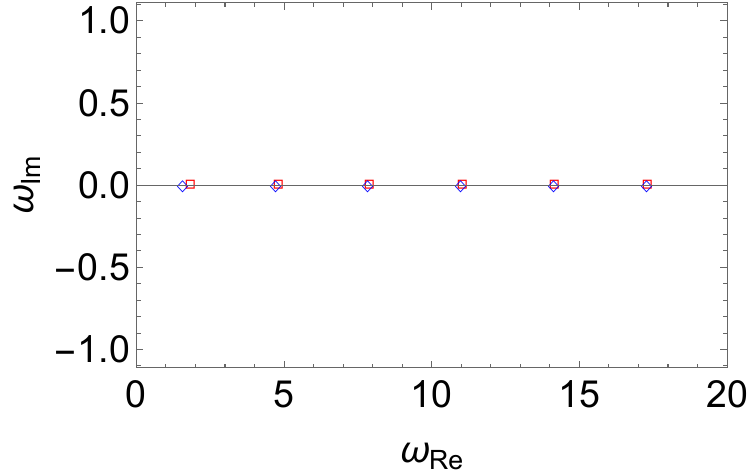}
\includegraphics[height=0.35\textwidth]{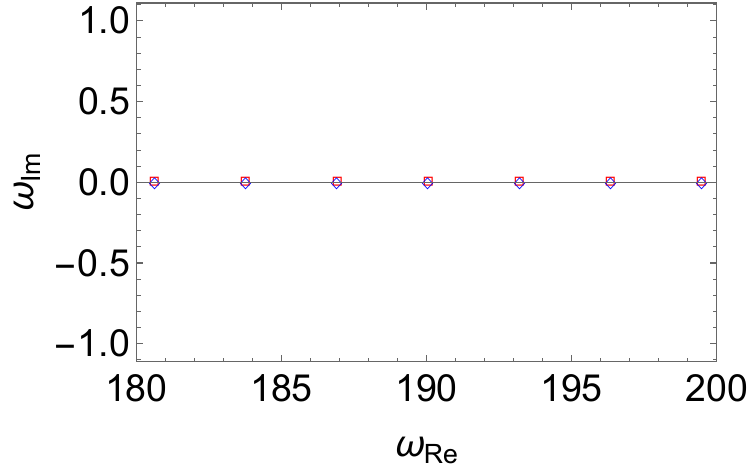}
}
\centerline{
\includegraphics[height=0.35\textwidth]{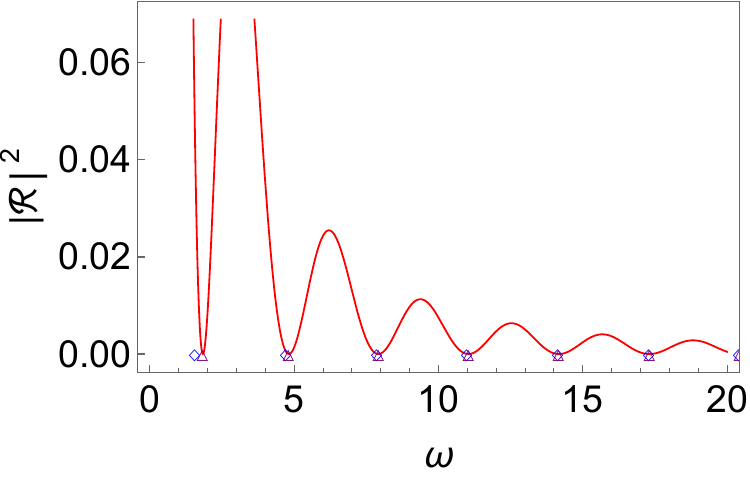}
\includegraphics[height=0.35\textwidth]{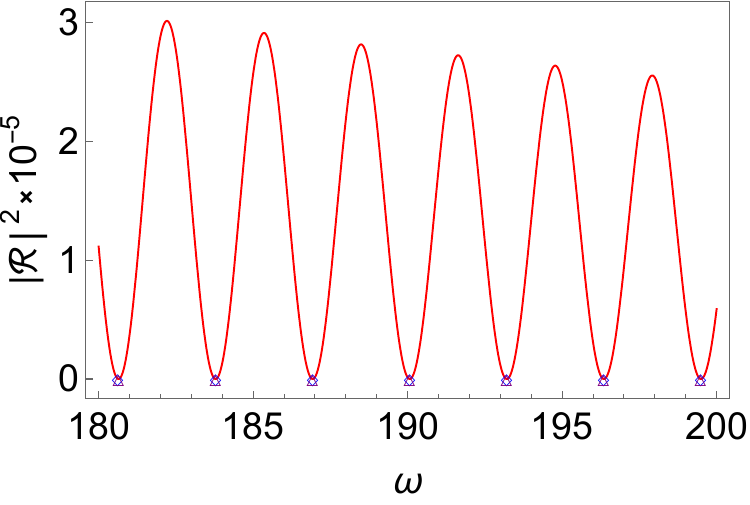}
}
\renewcommand{\figurename}{Fig.}
\vspace{-0.5cm}
\caption{The RSM and quasi-RSM modes and their asymptotic values for symmetric Damour-Solodukhin wormholes consisting of two identical delta effective potential barriers.
The calculations are carried out using the parameters $V_0^\mathrm{L}=V_0^\mathrm{R}=1$ and $x_c=\frac12$.
Upper row: The RSMs and quasi-RSMs coincide and are shown in empty red squares.
They are compared with the asymptotic values given by Eq.~\eqref{rootRSMread}, represented by empty blue diamonds.
Lower row: The reflection amplitude evaluated as a function of the frequency.
It vanishes identically at their local minima, which coincide with the RSMs.
The left column show the low-lying modes ($0\le \mathrm{Re}\omega\le 20$) and the right column illustrates the asymptotic modes ($180\le \mathrm{Re}\omega\le 200$).}
\label{fig_doubledelta_symmetric_DS}
\end{figure}

\begin{figure}[ht]
\centerline{
\includegraphics[height=0.35\textwidth]{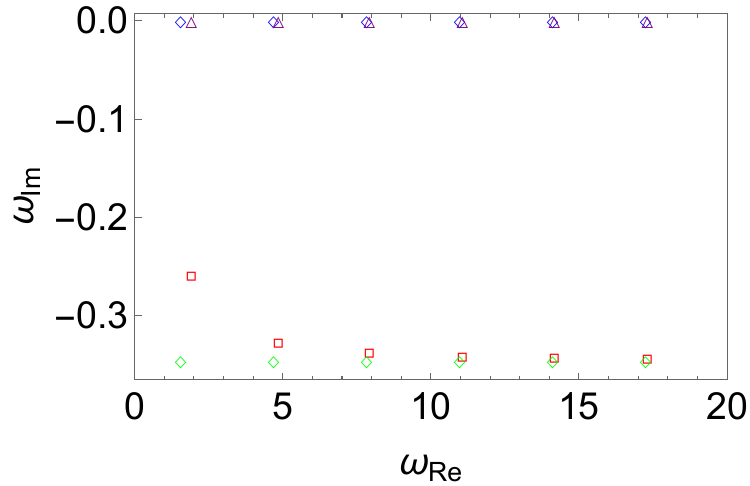}
\includegraphics[height=0.35\textwidth]{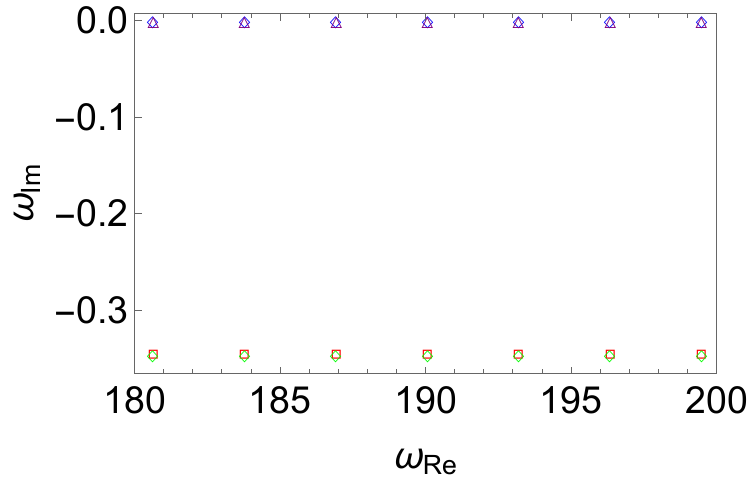}
}
\centerline{
\includegraphics[height=0.35\textwidth]{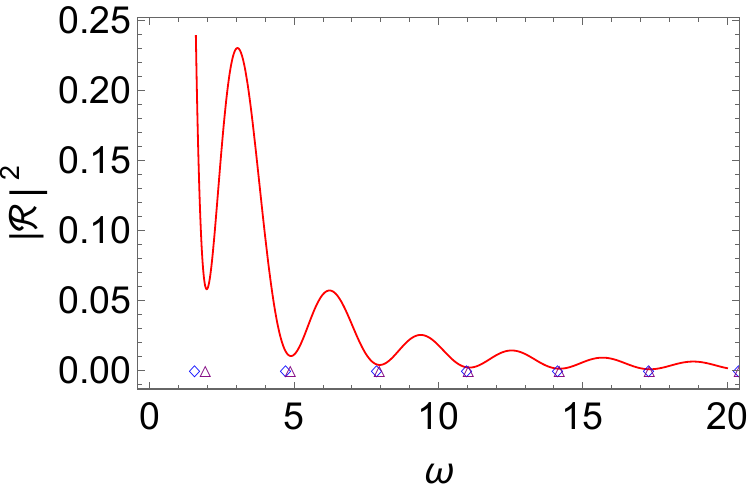}
\includegraphics[height=0.35\textwidth]{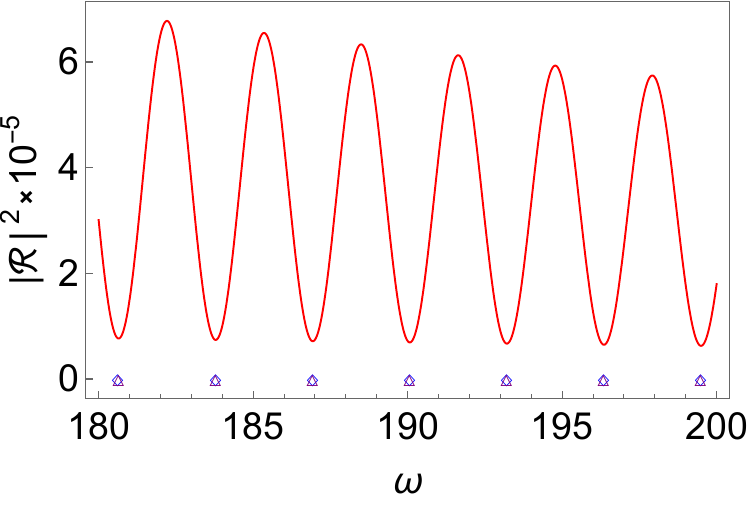}
}
\renewcommand{\figurename}{Fig.}
\vspace{-0.5cm}
\caption{The RSM and quasi-RSM modes and their asymptotic values for asymmetric Damour-Solodukhin wormholes consisting of two delta effective potential barriers with different magnitudes.
The calculations are carried out using the parameters $V_0^\mathrm{L}=2$, $V_0^\mathrm{R}=1$, and $x_c=\frac12$.
Upper row: The RSMs and quasi-RSMs are shown in empty red squares and empty purple triangles.
They are compared with the asymptotic values given by Eq.~\eqref{rootRSM}, represented by empty green diamonds, and their real parts, indicated by empty blue diamonds.
Lower row: The reflection amplitude evaluated as a function of the frequency.
It does not vanish at its local minima, which reflects that the RSMs are complex.
The left column show the low-lying modes ($0\le \mathrm{Re}\omega\le 20$) and the right column illustrates the asymptotic modes ($180\le \mathrm{Re}\omega\le 200$).}
\label{fig_doubledelta_asymmetric_DS}
\end{figure}

The numerical results are shown in Tab.~\ref{tabNumDblDelta}, Figs.~\ref{fig_doubledelta_symmetric_DS} and~\ref{fig_doubledelta_asymmetric_DS}.
As shown in the upper row of Fig.~\ref{fig_doubledelta_symmetric_DS}, the RSMs (empty red squares) are purely real in the case of symmetric Damour-Solodukhin wormholes.
This corresponds to the fact that the reflection amplitude vanishes identically at those values on the real frequency axis, as shown in the lower row of Fig.~\ref{fig_doubledelta_symmetric_DS}.
Also, the asymptotic result for RSMs (empty blue diamonds), Eq.~\eqref{rootRSMread}, approximates reasonably well the roots of Eq.~\eqref{EqRSM_DSdelta} even for the low-lying modes present in the left column of Fig.~\ref{fig_doubledelta_symmetric_DS}.

\begin{table}
\setlength{\tabcolsep}{12pt}
\caption{The RSMs and echo modes for Damour-Solodukhin wormholes consisting of two delta-function effective potential barriers.
We show both the numerical results and the estimations made at the limit $\omega\to \infty$.
For the symmetric case, one considers the metric parameters $V_0^\mathrm{L}=V_0^\mathrm{R}=1$ and $x_c=\frac12$.
For the asymmetric case, one considers the metric parameters $V_0^\mathrm{L}=2$, $V_0^\mathrm{R}=1$, and $x_c=\frac12$.} \label{tabNumDblDelta}
\begin{tabular}{c cccc cccc }
        \hline\hline
            RSMs                             & $n=0$ & 1 & 2 & 57 & 58 & 59 \\
          \hline
            Symmetric case                   & 1.8366 & 4.81584 & 7.91705 & 180.644 & 183.786 & 186.927  \\
            Est. by Eq.~\eqref{rootRSMread}  & 1.5708 & 4.71239 & 7.85398 & 180.642 & 183.783 & 186.925 \\
            Asymmetric case                   & 1.92285 & 4.86334  & 7.94756  & 180.646 & 183.787  & 186.929 \\
                                              &  -0.260979$i$ &  -0.328881$i$ &  -0.339671$i$ &  -0.34656$i$  &  -0.34656$i$ &  -0.346561$i$ \\
            Est. by Eq.~\eqref{rootRSM}       & 1.5708 & 4.71239 & 7.85398 & 180.642 & 183.783 & 186.925 \\
                                              &  -0.34657$i$ &  -0.34657$i$ &  -0.34657$i$ &  -0.34657$i$ &  -0.34657$i$ &  -0.34657$i$ \\
            \hline
            Echo modes                        & $n=0$ & 1 & 2 & 57 & 58 & 59 \\
          \hline
            Symmetric case                   & 1.21427 & 4.33175 & 7.56372  & 180.612 & 183.754  & 186.896   \\
                                             &  -0.95224$i$ &  -2.23338$i$ &  -2.75924$i$ &  -5.88994$i$ &  -5.90718  $i$ &  -5.92412$i$ \\
            Est. by Eq.~\eqref{exEstiQNM}   & 1.57080 & 4.71239 & 7.85398  & 180.642   & 183.783  & 186.925  \\
                                              & -1.14473$i$ &  -2.24334$i$ &  -2.75417$i$  &  -5.88966$i$ &  -5.9069$i$ &  -5.92385$i$ \\
            Asymmetric case                   & 1.54318 & 4.46629  & 7.64109    & 180.615   & 183.757  & 186.899  \\
                                              & -0.79374$i$ &  -1.8752$i$ &  -2.40346$i$  &  -5.54329$i$  &  -5.56053$i$ &  -5.57748$i$ \\
            Est. by Eq.~\eqref{exEstiQNM}     & 1.57080 & 4.71239 & 7.85398   & 180.642   & 183.783   & 186.925   \\
                                              & -0.79816$i$ &  -1.89677$i$ &  -2.40759$i$  &  -5.54309$i$ &  -5.56033$i$ &  -5.57728$i$ \\
         \hline\hline
\end{tabular}
\end{table}

For asymmetric Damour-Solodukhin wormholes, the RSMs (empty red squares) are complex and their real parts correspond to the quasi-RSMs (empty purple triangles) introduced in~\cite{agr-qnm-instability-63}, as shown in the upper row of Fig.~\ref{fig_doubledelta_asymmetric_DS}.
In this case, the reflection amplitude attains the local minima at the frequencies of quasi-RSMs on the real frequency axis, as shown in the lower row of the figure.
In such a more general context, the asymptotic expression for RSMs (empty green diamonds) given by Eq.~\eqref{rootRSM} provides a reasonable estimation of the asymptotic roots of Eq.~\eqref{EqRSM_DSdelta}, as shown in the bottom-right panel, while the deviations are apparent for low-lying modes, indicated in the bottom-left panel.
Their real parts (empty blue diamonds) closely follow the local minima of the reflection amplitude even for the low-lying modes present in the lower-left panel of Fig.~\ref{fig_doubledelta_asymmetric_DS}.
The numerical results are also presented in Tab.~\ref{tabNumDblDelta}.

For the QNMs, Eq.~\eqref{ConsymDS2} gives
\bqn
T^\mathrm{R}_{21}\left(T^\mathrm{L}\right)_{21}^{-1}\left[1 - e^{-4i\omega x_c}\frac{T^\mathrm{R}_{22}T^\mathrm{L}_{11}}{T^\mathrm{R}_{21}T^\mathrm{L}_{21}}\right]
=  
\frac{V_0^\mathrm{R}}{2\omega}\frac{V_0^\mathrm{L}}{2\omega} + e^{-4i\omega x_c}{\left(1+i\frac{V_0^\mathrm{R}}{2\omega}\right)\left(1+i\frac{V_0^\mathrm{L}}{2\omega}\right)} =  0 . \label{EqQNM_DSdelta}
\eqn
For asymptotic modes $\omega_n\to \infty$, the above equation simplifies to
\bqn
\frac{V_0^\mathrm{R}V_0^\mathrm{L}}{4\omega^2} + e^{-4i\omega x_c} = 0 ,
\eqn
which implies the following approximate form for asymptotic echo modes~\cite{agr-qnm-Poschl-Teller-16, agr-qnm-instability-65}
\bqn
\omega_n \simeq \left(n+\frac12\right)\frac{\pi}{2x_c} - i\frac{1}{2x_c}\left[\ln\left(\left(n+\frac12\right)\frac{\pi}{2x_c}\right)+\ln 2-\ln\sqrt{V_0^\mathrm{L}V_0^\mathrm{R}}\right] .\label{exEstiQNM}
\eqn
We note that the real parts of asymptotic QNMs governed by Eq.~\eqref{exEstiQNM} coincide with those of RSMs given by Eq.~\eqref{rootRSM}.
However, unlike RSMs, these QNMs typically possess non-vanishing imaginary parts, and they do not straightforwardly simplify to any case of particular interest.

\begin{figure}[ht]
\centerline{
\includegraphics[height=0.35\textwidth]{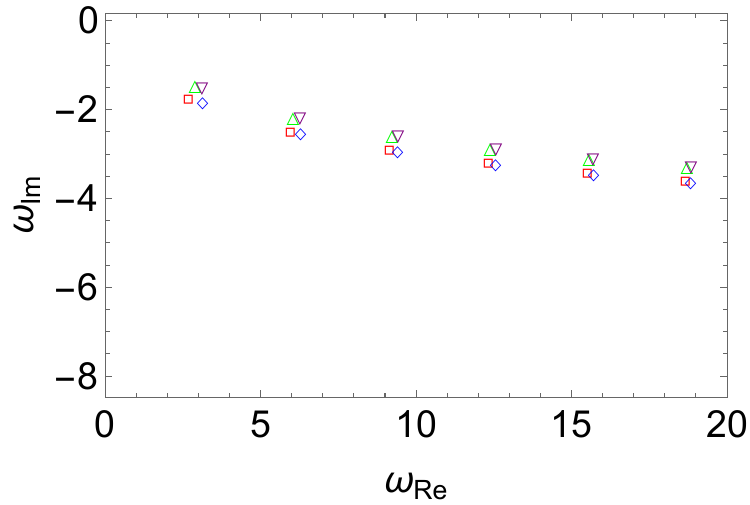}
\includegraphics[height=0.35\textwidth]{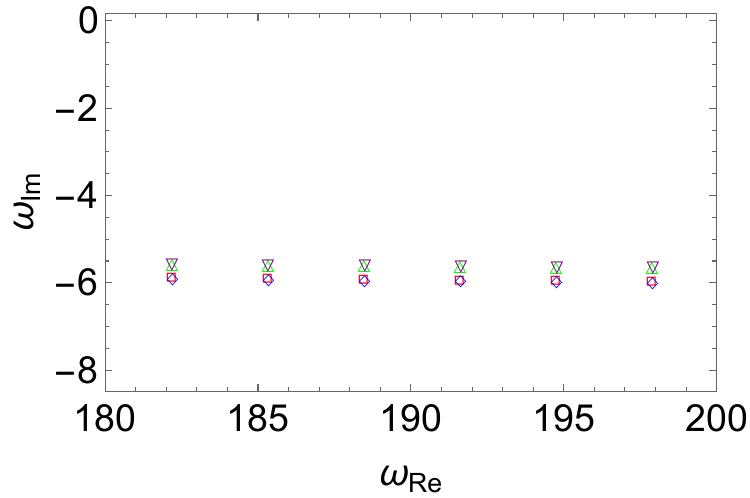}
}
\renewcommand{\figurename}{Fig.}
\vspace{-0.5cm}
\caption{The echo modes and their asymptotic values for symmetric and asymmetric Damour-Solodukhin wormholes consisting of two square barriers.
The calculations are carried out by using $x_c=\frac12$ and the parameters $V_0^\mathrm{L}=V_0^\mathrm{R}=1$ for symmetric case and $V_0^\mathrm{L}=2$, $V_0^\mathrm{R}=1$ for asymmetric case.
The numerical results for QNMs are shown in empty red squares (for symmetric wormhole) and empty green triangles (for asymmetric wormhole).
Those obtained by employing the estimated expression Eq.~\eqref{exEstiQNM} are indicated by empty blue diamonds (for symmetric wormhole) and empty purple inverted-triangles (for asymmetric wormhole).
The left column show the low-lying modes ($0\le \mathrm{Re}\omega\le 20$) and the right column illustrates the asymptotic modes ($180\le \mathrm{Re}\omega\le 200$).}
\label{fig_doubledelta_asymmetric_DS_QNM}
\end{figure}

The numerical results are shown in Tab.~\ref{tabNumDblDelta} and Fig.~\ref{fig_doubledelta_asymmetric_DS_QNM}.
Unlike RSMs, for both symmetric and asymmetric echo modes, the imaginary parts of the frequencies do not vanish.
Specifically, the difference between symmetric and asymmetric Damour-Solodukhin wormholes is not significant.
This is because the term $\ln\sqrt{V_0^\mathrm{L}V_0^\mathrm{R}}$ only furnishes a minor correction to the primary contribution associated with the order $n$, which only depends on the length scale $x_c$.
One observes that the expression Eq.~\eqref{exEstiQNM} provides a satisfactory estimation of the asymptotic modes, particularly for the high overtones shown in the right panel, where $|\mathrm{Re}\omega| \gg |\mathrm{Im}\omega|$.
We note that the quasinormal frequencies from the two delta function barriers, which are both constituted by a purely imaginary mode $\omega^\mathrm{L, R}_\mathrm{BH}=-iV^\mathrm{L, R}_0/2$, did not appear as a part of QNMs, nor RSMs.
The numerical results are also given in Tab.~\ref{tabNumDblDelta}.

As the second example, we consider the double-square potential barrier.
\begin{equation}
V_\mathrm{BH}^\mathrm{L,R}(x) =
\begin{cases}
 \  0 &  x \le -\frac12 W^\mathrm{L,R}, \\
 \  H^\mathrm{L, R} &  -\frac12 W^\mathrm{L,R} < x \le \frac12 W^\mathrm{L,R}, \\
 \  0 &  x > \frac12 W^\mathrm{L,R} , 
\end{cases}
\label{VsquareCase}
\end{equation}
The corresponding transfer matrices read
\bqn
T 
=
\begin{pmatrix}{T}_{11}&{T}_{12}\\ {T}_{21}&{T}_{22}\end{pmatrix} , \label{TsquareSimplified}
\eqn
where
\bqn
{T}_{11}&=& \frac12 e^{-iW^\mathrm{L,R}\omega}\left[2\cos\left(W^\mathrm{L,R}\sqrt{\omega^2-V^\mathrm{L,R}}\right)- i\frac{\left(V^\mathrm{L,R}-2\omega^2\right)\sin\left(W^\mathrm{L,R}\sqrt{\omega^2-V^\mathrm{L,R}}\right)}{\omega\sqrt{\omega^2-V^\mathrm{L,R}}}\right],\nb\\
{T}_{12}&=& -i\frac12\frac{V^\mathrm{L,R}\sin\left(W^\mathrm{L,R}\sqrt{\omega^2-V^\mathrm{L,R}}\right)}{\omega\sqrt{\omega^2-V^\mathrm{L,R}}},\nb\\
{T}_{21}&=& i\frac12\frac{V^\mathrm{L,R}\sin\left(W^\mathrm{L,R}\sqrt{\omega^2-V^\mathrm{L,R}}\right)}{\omega\sqrt{\omega^2-V^\mathrm{L,R}}} ,\nb\\
{T}_{22}&=& \frac12 e^{iW^\mathrm{L,R}\omega}\left[2\cos\left(W^\mathrm{L,R}\sqrt{\omega^2-V^\mathrm{L,R}}\right)+ i\frac{\left(V^\mathrm{L,R}-2\omega^2\right)\sin\left(W^\mathrm{L,R}\sqrt{\omega^2-V^\mathrm{L,R}}\right)}{\omega\sqrt{\omega^2-V^\mathrm{L,R}}}\right] ,
\eqn
and
\bqn
T^{-1} 
=
\begin{pmatrix}\left(T^{-1}\right)_{11}&\left(T^{-1}\right)_{12}\\ \left(T^{-1}\right)_{21}&\left(T^{-1}\right)_{22}\end{pmatrix} , \label{TsquareInvSimplified}  
\eqn
where
\bqn
\left(T^{-1}\right)_{11}&=& \frac12 e^{iW^\mathrm{L,R}\omega}\left[2\cos\left(W^\mathrm{L,R}\sqrt{\omega^2-V^\mathrm{L,R}}\right)+ i\frac{\left(V^\mathrm{L,R}-2\omega^2\right)\sin\left(W^\mathrm{L,R}\sqrt{\omega^2-V^\mathrm{L,R}}\right)}{\omega\sqrt{\omega^2-V^\mathrm{L,R}}}\right],\nb\\
\left(T^{-1}\right)_{12}&=& i\frac12\frac{V^\mathrm{L,R}\sin\left(W^\mathrm{L,R}\sqrt{\omega^2-V^\mathrm{L,R}}\right)}{\omega\sqrt{\omega^2-V^\mathrm{L,R}}},\nb\\
\left(T^{-1}\right)_{21}&=& -i\frac12\frac{V^\mathrm{L,R}\sin\left(W^\mathrm{L,R}\sqrt{\omega^2-V^\mathrm{L,R}}\right)}{\omega\sqrt{\omega^2-V^\mathrm{L,R}}} ,\nb\\
\left(T^{-1}\right)_{22}&=& \frac12 e^{-iW^\mathrm{L,R}\omega}\left[2\cos\left(W^\mathrm{L,R}\sqrt{\omega^2-V^\mathrm{L,R}}\right)- i\frac{\left(V^\mathrm{L,R}-2\omega^2\right)\sin\left(W^\mathrm{L,R}\sqrt{\omega^2-V^\mathrm{L,R}}\right)}{\omega\sqrt{\omega^2-V^\mathrm{L,R}}}\right] .
\eqn
We relegate some specific details to Appx.~\ref{appD}.

For the RSMs, we find
%\bqn T^\mathrm{R}_{11}\left(T^\mathrm{L}\right)_{21}^{-1}\left[1 - e^{-4i\omega x_c}\frac{T^\mathrm{R}_{12}}{T^\mathrm{L}_{12}}\frac{T^\mathrm{L}_{22}}{T^\mathrm{R}_{11}}\frac{\left(T^\mathrm{L}\right)_{12}^{-1}}{\left(T^\mathrm{L}\right)_{21}^{-1}}\right] = 0 \eqn
\bqn
0~&&= 1 + e^{-4i\omega x_c}\nb\\
&&\times\frac{V^\mathrm{R}\sin\left(W^\mathrm{R}\sqrt{\omega^2-V^\mathrm{R}}\right)}{V^\mathrm{L}\sin\left(W^\mathrm{L}\sqrt{\omega^2-V^\mathrm{L}}\right)}\frac{\sqrt{\omega^2-V^\mathrm{L}}}{\sqrt{\omega^2-V^\mathrm{R}}} \nb\\
&&\times e^{i\left(W^\mathrm{L}+W^\mathrm{R}\right)\omega}\frac{2\cos\left(W^\mathrm{L}\sqrt{\omega^2-V^\mathrm{L}}\right)+ i\frac{\left(V^\mathrm{L}-2\omega^2\right)\sin\left(W^\mathrm{L}\sqrt{\omega^2-V^\mathrm{L}}\right)}{\omega\sqrt{\omega^2-V^\mathrm{L}}}}{2\cos\left(W^\mathrm{R}\sqrt{\omega^2-V^\mathrm{R}}\right)- i\frac{\left(V^\mathrm{R}-2\omega^2\right)\sin\left(W^\mathrm{R}\sqrt{\omega^2-V^\mathrm{R}}\right)}{\omega\sqrt{\omega^2-V^\mathrm{R}}}}  , \label{EqRSM_DS_square}
\eqn
by substituting Eqs.~\eqref{TsquareSimplified} and~\eqref{TsquareInvSimplified} into Eq.~\eqref{ConsymDS}.

For asymptotical RSMs $\omega_n\to \infty$ in Damour-Solodukhin wormholes, the third line of the above equation approaches 1.
However, the second line, particularly the ratio between two sine functions, does not have a well-defined limit.
In fact, one observes that each sine function contains the factor $1-e^{-2i\omega W^\mathrm{L, R}}$, which oscillates constantly as $\mathrm{Re}\omega\to \infty$.
In other words, each one of the two square potential barriers brings its intrinsic echo modes, related to its length scale $W^\mathrm{L, R}$.
However, as one typically has $W^\mathrm{L, R} \ll x_c$, these modes are characterized by a larger interval in the frequency domain, to the extent that the internal scale of the barrier becomes physically irrelevant.
%Indeed, at the limit $W^\mathrm{L, R}\to 0$, it is immediately recognized that the ratio falls back to Eq.~\eqref{asyDeltaRSM}, given that $W^\mathrm{L, R}V^\mathrm{L, R} \to V_0^\mathrm{L, R}$.
Specifically, as one is interested in successive RSMs or echo modes at the limit $\omega\to\infty$, the ratio between the two sine functions furnishes a correction to the RSMs which has a period inversely proportional to the width of the barrier.
When $\omega W^\mathrm{L, R} \ll x_c$, the asymptotic RSMs read
\bqn
\omega_n \simeq \left(n+\frac12\right)\frac{\pi}{2x_c} - i \frac{1}{4x_c} \ln\left(\frac{V^\mathrm{L}}{V^\mathrm{R}}\right) ,\label{rootRSMsquare}
\eqn
which again falls back to purely real ones Eq.~\eqref{rootRSMread} for symmetric Damour-Solodukhin wormholes.

For the QNMs, Eq.~\eqref{ConsymDS2} gives
%\bqn T^\mathrm{R}_{12}\left(T^\mathrm{L}\right)_{21}^{-1}\left[1 - e^{-4i\omega x_c}\frac{T^\mathrm{R}_{22}T^\mathrm{L}_{22}}{T^\mathrm{R}_{21}T^\mathrm{L}_{21}}\right] =  \eqn 
\bqn
0 &=& 1 + e^{-4i\omega x_c} \nb\\
&\times& {e^{iW^\mathrm{R}\omega}\left[2\cos\left(W^\mathrm{R}\sqrt{\omega^2-V^\mathrm{R}}\right)+ i\frac{\left(V^\mathrm{R}-2\omega^2\right)\sin\left(W^\mathrm{R}\sqrt{\omega^2-V^\mathrm{R}}\right)}{\omega\sqrt{\omega^2-V^\mathrm{R}}}\right]}\nb\\
&\times& {e^{iW^\mathrm{L}\omega}\left[2\cos\left(W^\mathrm{L}\sqrt{\omega^2-V^\mathrm{L}}\right)+ i\frac{\left(V^\mathrm{L}-2\omega^2\right)\sin\left(W^\mathrm{L}\sqrt{\omega^2-V^\mathrm{L}}\right)}{\omega\sqrt{\omega^2-V^\mathrm{L}}}\right] } \nb\\
&\times& \frac{\omega\sqrt{\omega^2-V^\mathrm{R}}}{V^\mathrm{R}\sin\left(W^\mathrm{R}\sqrt{\omega^2-V^\mathrm{R}}\right)}
\frac{\omega\sqrt{\omega^2-V^\mathrm{L}}}{V^\mathrm{L}\sin\left(W^\mathrm{L}\sqrt{\omega^2-V^\mathrm{L}}\right)}, \label{EqQNM_DSsquare}
\eqn
which implies the following approximate form for asymptotic echo modes
\bqn
\omega_n \simeq \left(n+\frac12\right)\frac{\pi}{2x_c} - i\frac{1}{2x_c}\left[2\ln\left(\left(n+\frac12\right)\frac{\pi}{2x_c}\right)+\ln 2-\ln\sqrt{V^\mathrm{L}V^\mathrm{R}}\right] .\label{exEstiQNMsquare}
\eqn
Again, for asymptotic modes, the real parts of Eq.~\eqref{exEstiQNMsquare} coincide with those of RSMs given by Eq.~\eqref{rootRSMsquare}.

\begin{figure}[ht]
\centerline{
\hspace{-1.5cm}
\includegraphics[height=0.35\textwidth]{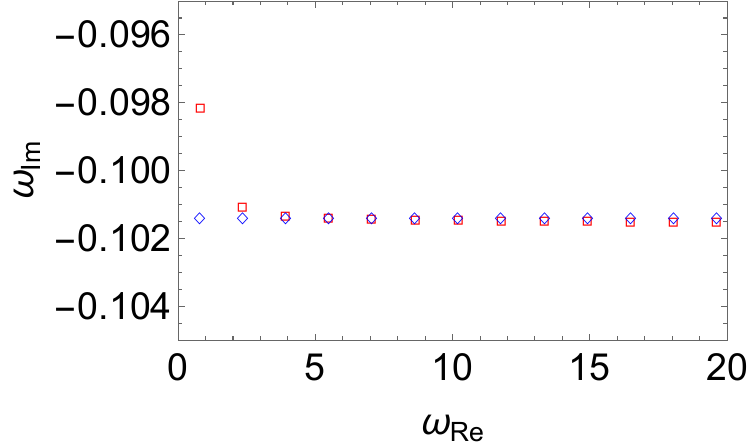}
\includegraphics[height=0.35\textwidth]{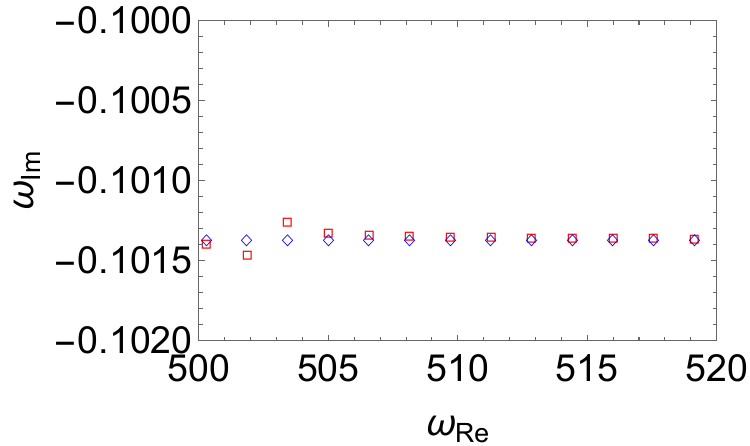}
}
\centerline{
\includegraphics[height=0.33\textwidth]{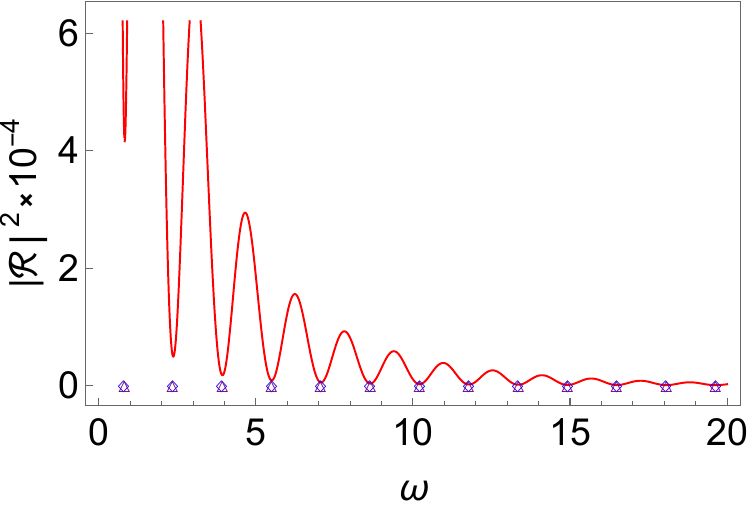}
\hspace{1.3cm}
\includegraphics[height=0.33\textwidth]{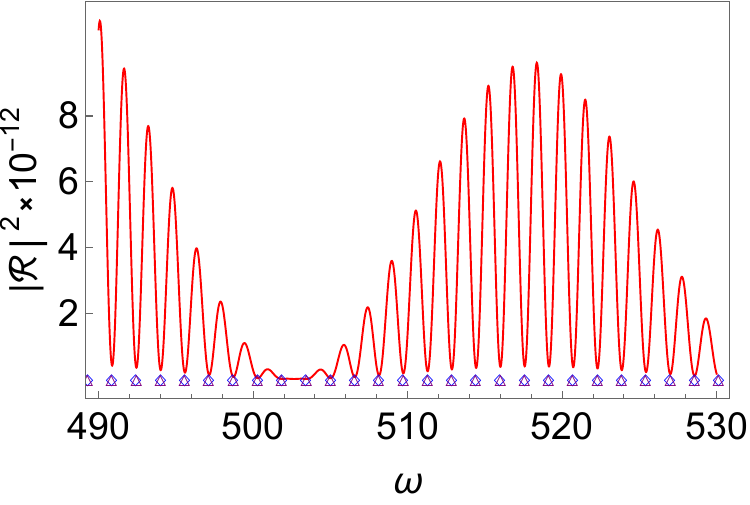}
}
\centerline{
\hspace{-1.5cm}
\includegraphics[height=0.35\textwidth]{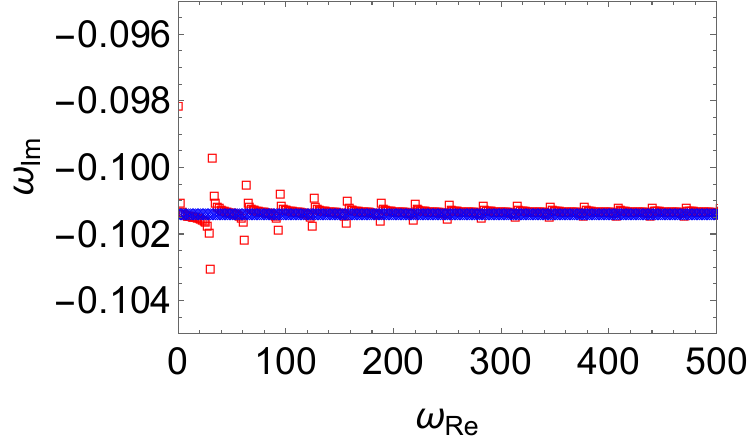}
\includegraphics[height=0.35\textwidth]{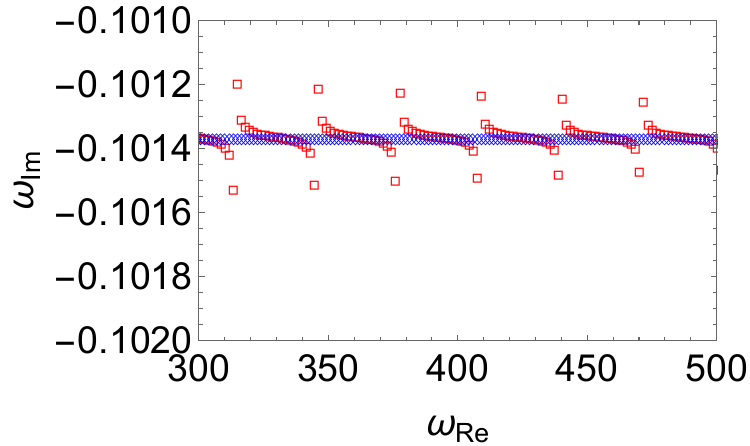}
}
\renewcommand{\figurename}{Fig.}
\vspace{-0.5cm}
\caption{The RSM and quasi-RSM modes and their asymptotic values for asymmetric Damour-Solodukhin wormholes consisting of two square barriers with different magnitudes.
The calculations are carried out using the parameters $V^\mathrm{L}=1$, $V^\mathrm{R}=\frac23$, $W^\mathrm{L}=W^\mathrm{R}=\frac{1}{10}$, and $x_c=1$.
Top row: The RSMs are shown in empty red squares, and they are compared with the asymptotic values given by Eq.~\eqref{rootRSM}, represented by empty blue diamonds.
Middle row: The reflection amplitude evaluated as a function of the frequency.
It does not vanish at their local minima, which reflects that the RSMs are complex.
The quasi-RSMs and the real parts of the RSMs are represented by empty blue diamonds and empty purple triangles.
The left column show the low-lying modes ($0\le \mathrm{Re}\omega\le 20$) and the right column illustrates the asymptotic modes ($180\le \mathrm{Re}\omega\le 200$).
Bottom row: An overhead shot of RSMs for the asymmetric wormhole.
The deviations from the estimation Eq.~\eqref{rootRSM} decrease with increasing frequency, and are observed to have a period of $10\pi$, which reflects the finite size of the square barriers. 
}
\label{fig_doublesquare_symmetric_DS}
\end{figure}

\begin{figure}[ht]
\centerline{
\includegraphics[height=0.35\textwidth]{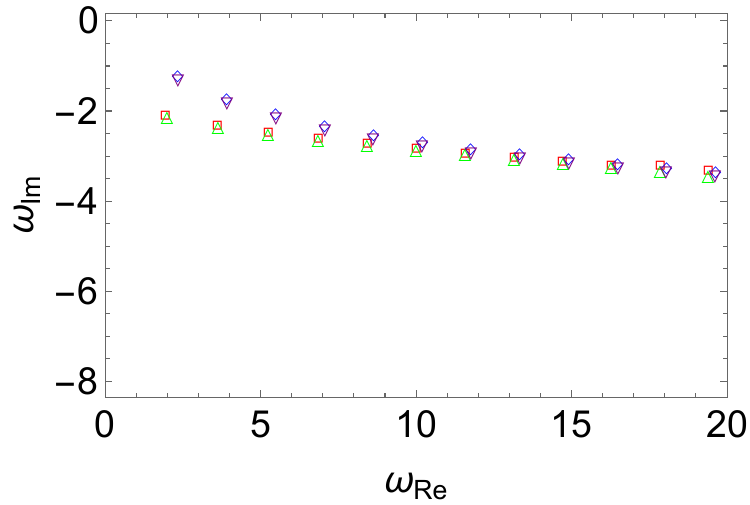}
\includegraphics[height=0.35\textwidth]{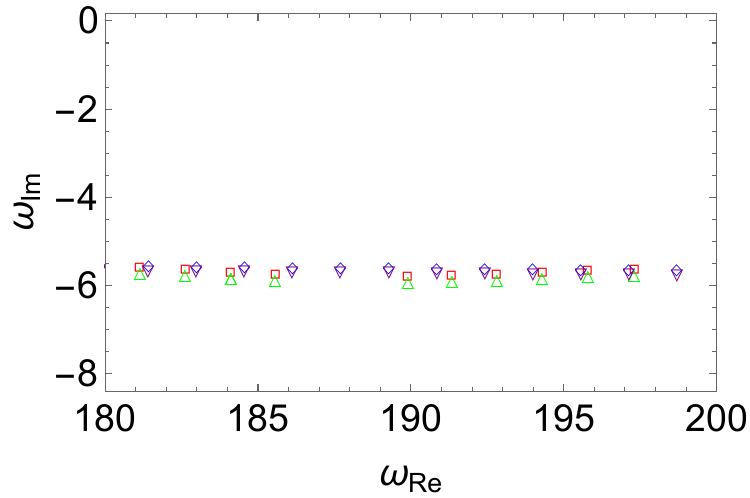}
}
\centerline{
\includegraphics[height=0.35\textwidth]{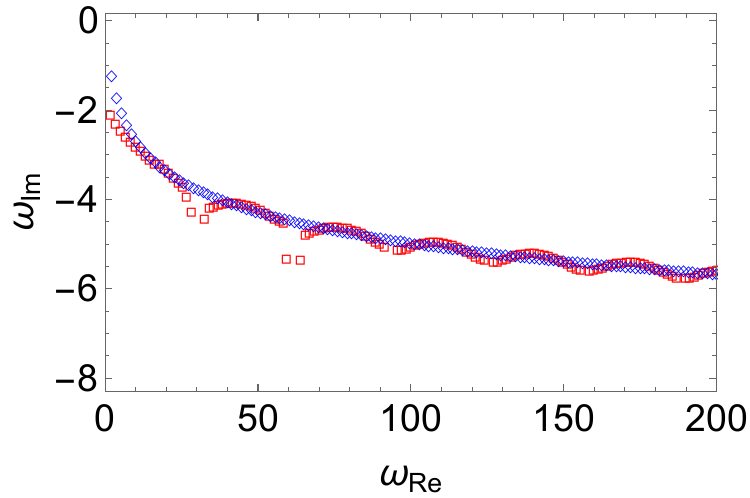}
\includegraphics[height=0.35\textwidth]{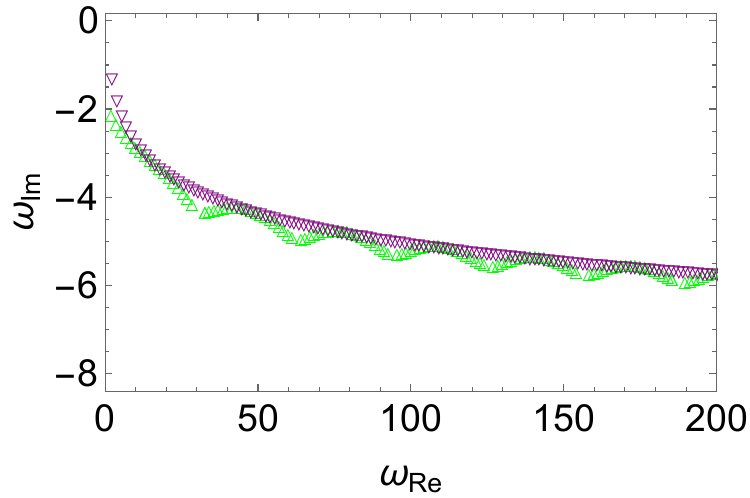}
}
\renewcommand{\figurename}{Fig.}
\vspace{-0.5cm}
\caption{The echo modes and their asymptotic values for symmetric and asymmetric Damour-Solodukhin wormholes consisting of two square effective potential barriers.
The calculations are carried out by using $W^\mathrm{L}=W^\mathrm{R}=\frac{1}{10}$, $x_c=1$, and the parameters $V^\mathrm{L}=V^\mathrm{R}=1$ for symmetric case and $V^\mathrm{L}=1$, $V^\mathrm{R}=\frac23$ for asymmetric case.
The numerical results for QNMs are shown in empty red squares (for symmetric wormhole) and empty green triangles (for asymmetric wormhole).
Those obtained by employing the estimated expression Eq.~\eqref{exEstiQNM} are indicated by empty blue diamonds (for symmetric wormhole) and empty purple inverted-triangles (for asymmetric wormhole).
Top row: A close-up view of the low-lying ($0\le \mathrm{Re}\omega\le 20$) and asymptotic modes ($180\le \mathrm{Re}\omega\le 200$).
Bottom row: An overhead shot of echo modes in the symmetric (bottom-left) and asymmetric (bottom-right) wormholes, compared with estimations.
Similar to the RSMs shown in Fig.~\ref{fig_doublesquare_symmetric_DS}, an overall modulation of the period $10\pi$ is observed, which can be attributed to the width of the square barriers.}
\label{fig_doublesquare_asymmetric_DS_QNM}
\end{figure}

The numerical results are shown in Tab.~\ref{tabNumDblSquare} and Figs.~\ref{fig_doublesquare_symmetric_DS} and~\ref{fig_doublesquare_asymmetric_DS_QNM}.
In Fig.~\ref{fig_doubledelta_asymmetric_DS}, one presents the resulting RSMs for asymmetric Damour-Solodukhin wormholes (empty red squares), and their real parts, which correspond to the quasi-RSMs (empty purple triangles). 
As shown in the upper row, Eq.~\eqref{rootRSMsquare} (empty blue diamonds) provides a reasonable estimation.
The reflection amplitude attains the local minima at the frequencies of quasi-RSMs on the real frequency axis, as shown in the middle row.
The real parts of Eq.~\eqref{rootRSMsquare} (empty blue diamonds) closely follow the local minima of the reflection amplitude.
Regarding the entire RSM spectrum, one observes that the deviations from Eq.~\eqref{rootRSMsquare} decrease with increasing frequency and are characterized by a period of $\pi/W^\mathrm{L, R}$.
It is understood that the latter reflects the internal scale of the potential barriers, but it is a minor feature when compared to the RSMs as shown in the first row of Fig.~\ref{fig_doublesquare_symmetric_DS}.
The numerical values of the RSMs are also presented in Tab.~\ref{tabNumDblSquare}.

The results of the echo modes are presented in Fig.~\ref{fig_doublesquare_asymmetric_DS_QNM}.
Unlike RSMs, for both symmetric and asymmetric echo modes, the imaginary parts of the frequencies do not vanish.
Also, the difference between symmetric and asymmetric Damour-Solodukhin wormholes is not significant, as the primary contribution comes from the terms associated with the order $n$, which depend exclusively on the length scale $x_c$.
Except for the few low-lying modes that deviate more significantly, the approximation Eq.~\eqref{exEstiQNMsquare} provides a satisfactory estimation of the asymptotic modes, particularly for the high overtones shown in the top-right panel, where $|\mathrm{Re}\omega| \gg |\mathrm{Im}\omega|$.
Again, when a modulation is observed in the spectrum, as shown in the bottom row of Fig.~\ref{fig_doublesquare_asymmetric_DS_QNM}.
The modulation has the same period observed for RSMs, which is governed by the size of the potential barriers.

Before closing this subsection, it is instructive to directly compare the RSM spectrum with that of the QNMs. 
From the examples elaborated above, it is observed that the asymptotic modes in these cases have identical spacing along the real frequency axis, as analytically assessed in the previous sections. 
However, the RSM spectrum lies closer to the real axis than the QNMs, as demonstrated in Tabs.~\ref{tabNumDblDelta} and~\ref{tabNumDblSquare}. 
While this feature guarantees identical echo periods in both scenarios, it leads to a nontrivial difference in the resulting waveforms. 
As will be elaborated in the following subsection, for a given identical source, the waveforms associated with RSMs are more pronounced than their counterparts related to QNMs.

\begin{table}
\setlength{\tabcolsep}{12pt}
\caption{The RSMs and echo modes for Damour-Solodukhin wormholes consisting of two square barriers.
We show both the numerical results and the estimations made at the limit $\omega\to \infty$.
The calculations are carried out by using $W^\mathrm{L}=W^\mathrm{R}=\frac{1}{10}$, $x_c=1$, and the parameters $V^\mathrm{L}=V^\mathrm{R}=1$ for symmetric case and $V^\mathrm{L}=1$, $V^\mathrm{R}=\frac23$ for asymmetric case.} \label{tabNumDblSquare}
\begin{tabular}{c cccc cccc }
        \hline\hline
            RSMs                             & $n=0$ & 1 & 2 & 318 & 319 & 320 \\
          \hline
            Symmetric case                   & 2.36677 & 3.93336 & 5.50234 & 500.299 & 501.869 & 503.44 \\
            Est. by Eq.~\eqref{rootRSMsquare}  & 2.35619 & 3.92699 & 5.49779 & 500.299 & 501.869 & 503.44 \\
            Asymmetric case                  & 2.365 & 3.93229 & 5.50158 & 500.299 & 501.869 & 503.44 \\
                                              &  -0.101099$i$ &  -0.101359$i$ &  -0.101433$i$ &  -0.101401$i$ &  -0.10147$i$ &  -0.101263$i$ \\
            Est. by Eq.~\eqref{rootRSMsquare}      & 2.35619 & 3.92699  & 5.49779  & 500.299  & 501.869   & 503.44 \\
                                              &  -0.101366$i$ &  -0.101366$i$ &  -0.101366$i$ &  -0.101366$i$  &  -0.101366$i$ &  -0.101366$i$ \\
            \hline
            Echo modes                        & $n=0$ & 1 & 2 & 57 & 58 & 59 \\
          \hline
            Symmetric case                   & 1.95716  & 3.64419  & 5.26738  & 181.141  & 182.641 & 184.121   \\
                                              &  -2.00569$i$ &  -2.22675$i$ &  -2.39116$i$ &  -5.60322$i$ &  -5.65599$i$ &  -5.70883$i$ \\
            Est. by Eq.~\eqref{exEstiQNMsquare}  & 2.35619 & 3.92699  & 5.49779 & 181.427   & 182.998   & 184.569  \\
                                              &  -1.20362$i$ &  -1.71445$i$ &  -2.05092$i$ &  -5.54743$i$ &  -5.55605$i$ &  -5.56459$i$ \\
            Asymmetric case                   & 1.93933 & 3.63131  & 5.25729  & 181.036 & 182.639 & 183.868   \\
                                              &  -2.1203$i$ &  -2.33353$i$ &  -2.49537$i$ &  -5.89956$i$ &  -6.11234$i$ &  -6.23241$i$ \\
            Est. by Eq.~\eqref{exEstiQNMsquare} & 2.35619 & 3.92699  & 5.49779 & 181.137   & 182.635   & 184.117    \\
                                              &  -1.30499$i$ &  -1.81581$i$ &  -2.15229$i$ &  -5.69963$i$ &  -5.75401$i$ &  -5.81016$i$ \\
         \hline\hline
\end{tabular}
\end{table}

\subsection{The waveforms of RSMs and echo modes}\label{sec5.2}

We now elaborate on the similarities and differences between the RSMs and echo modes through their time-domain waveforms.  
In this subsection, the waveforms are computed using Green's functions constructed using Eqs.~\eqref{FormalGreen} and~\eqref{FormalGreenTilde}.  
The primary difference between them arises from the distinct boundary conditions imposed on $f(\omega, x)$ and $\widetilde{f}(\omega, x)$.  

We consider asymmetric Damour-Solodukhin wormholes composed of two delta-function effective potential barriers with unequal magnitudes, adopting the same parameters as in the preceding subsection for the spectral calculations.

\begin{figure}[ht]
\centerline{
\includegraphics[height=0.35\textwidth]{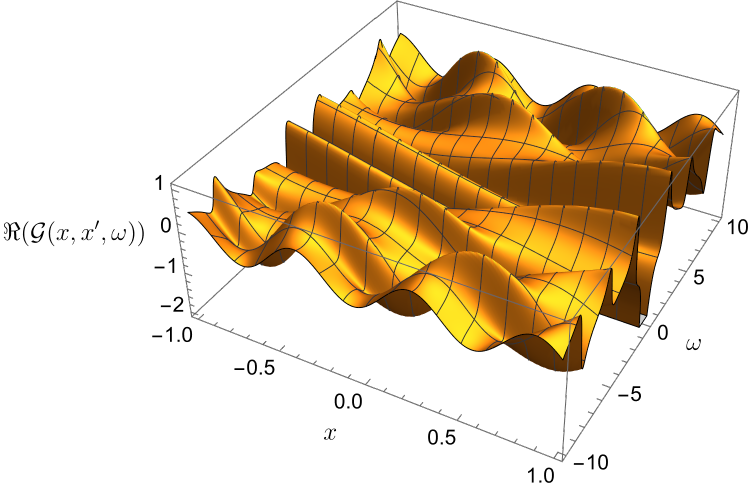}
\includegraphics[height=0.35\textwidth]{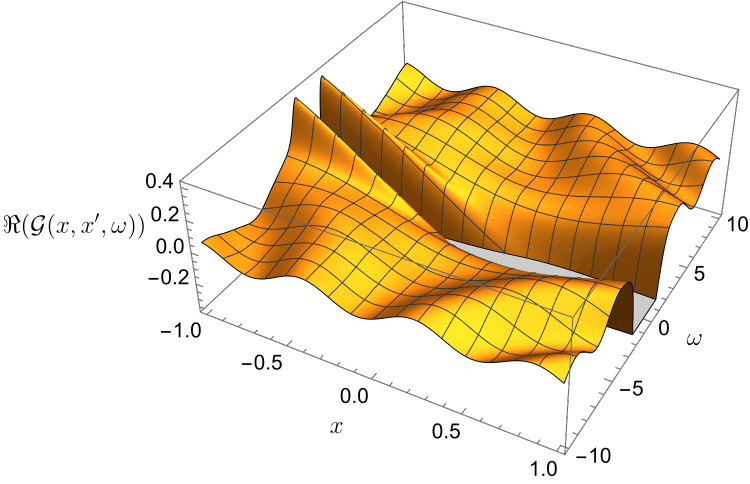}
}
\centerline{
\includegraphics[height=0.35\textwidth]{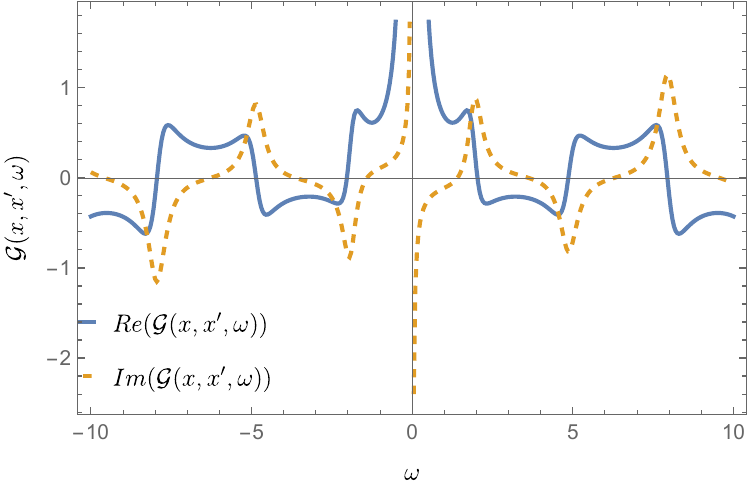}
\includegraphics[height=0.35\textwidth]{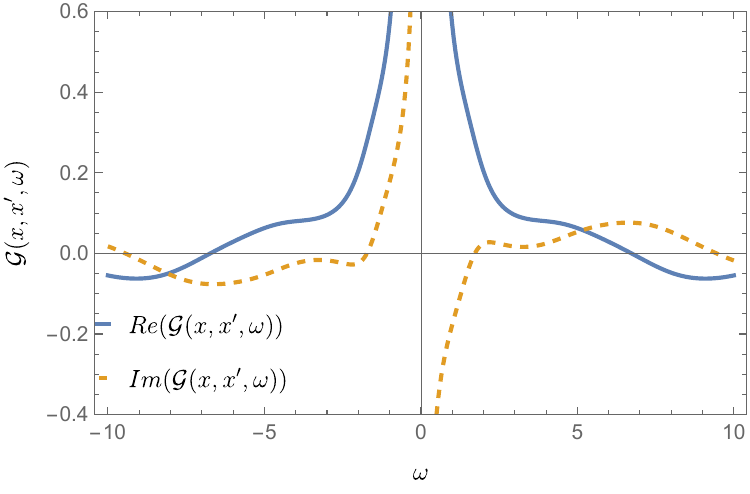}
}
\renewcommand{\figurename}{Fig.}
\vspace{-0.5cm}
\caption{The frequency-domain Green's functions, Eqs.~\eqref{FormalGreen} and~\eqref{FormalGreenTilde}, for echo modes and RSMs of asymmetric Damour-Solodukhin wormholes composed of two delta-function effective potential barriers with different magnitudes.  
The calculations are performed using the same parameters as in Fig.~\ref{fig_doubledelta_asymmetric_DS}, with $V_0^\mathrm{L}=2$, $V_0^\mathrm{R}=1$, and $x_c=\frac12$.  
Upper row: The real part of the Green's functions as a function of $\omega$ and $x$, with $x'=y_0=-\frac14$.  
Bottom row: The real and imaginary parts of the Green's functions on the spatial slice $x=\frac14$.
The results associated with the RSMs are shown in the left column, while those corresponding to the echo modes are displayed in the right column.}
\label{fig_doublesquare_asymmetric_GF}
\end{figure}

Using the explicit forms Eqs.~\eqref{G2delta} and~\eqref{G2deltaTilde} derived in Appx.~\ref{appE}, the numerical results for the Green's functions are shown in Fig.~\ref{fig_doublesquare_asymmetric_GF}.  
It is apparent that the two Green's functions differ significantly from each other.
Meanwhile, unlike those observed with the Regge-Wheeler potential (cf. Fig.~1 of~\cite{agr-strong-lensing-correlator-15}), they both exhibit strong oscillations in the spatial and frequency domains.  
This behavior stems from the poles lying close to the real frequency axis, as demonstrated in the preceding subsection.

We now compute the waveforms.  
To enable a meaningful comparison between the waveforms for RSMs and echo modes, for both cases, we adopt the following identical source in the frequency domain:
\bqn
S(\omega, y) = \sigma_0\delta(y-y_0)\frac{\omega^2}{(\omega-\omega_0)^3} ,\label{pulseS}
\eqn
where $\sigma_0=1$, $y_0=-\frac14$, and $\omega_0=-4+2i$.  
This represents a pulse localized at $y = y_0$, with a spread in the frequency domain.  
Physically, this source term encodes the information about the initial/boundary conditions from the Fourier~\cite{agr-qnm-tail-06} and Laplace~\cite{agr-qnm-29, agr-qnm-review-02} transform perspectives.  
We also note that the specific form of Eq.~\eqref{pulseS} is not essential.
As detailed further in Appx.~\ref{appE}, the numerator $\omega^2$ suppresses the pole at the origin, the denominator ensures faster convergence of the inverse Fourier transform, and the additional pole at $\omega=\omega_0$ plays no role since the contour integration follows a large circle in the lower half of the complex frequency plane.  
The resulting frequency-domain waveform is given by
\bqn
\Psi(\omega, x)=\int_y G(\omega, x, y)S(\omega, y) dy = \sigma_0 G(\omega, x, y_0)\frac{\omega^2}{(\omega-\omega_0)^3} ,\label{omegaPsi}
\eqn
which can be transformed into the time domain via the inverse Fourier transform. 

\begin{figure}[ht]
\centerline{
\includegraphics[height=0.35\textwidth]{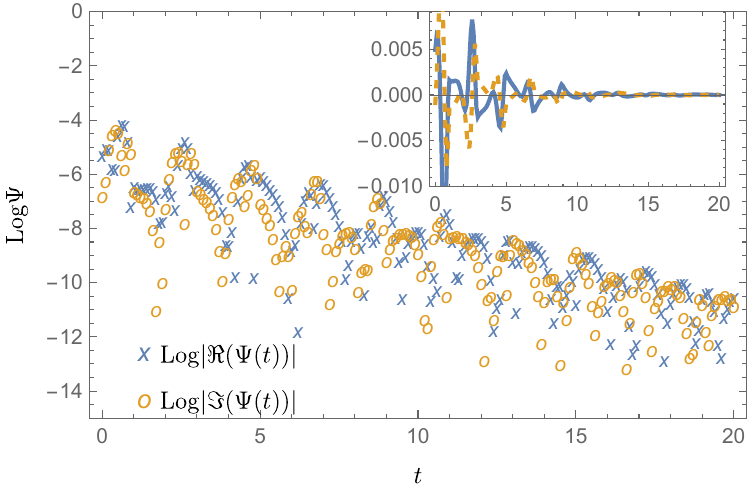}
\includegraphics[height=0.35\textwidth]{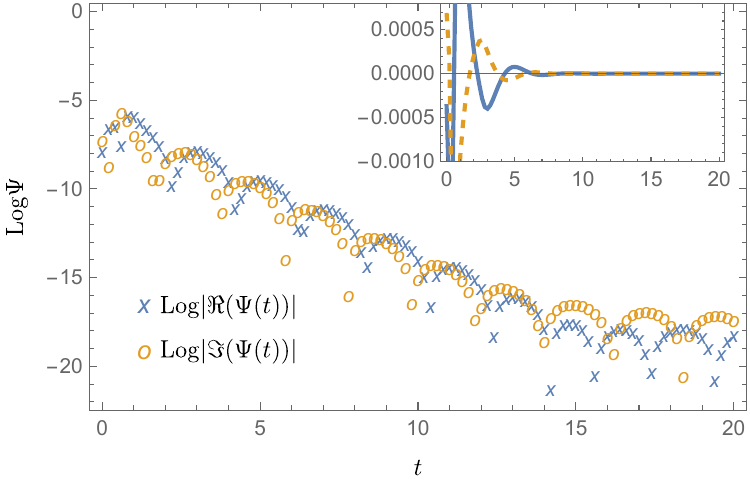}
}
\renewcommand{\figurename}{Fig.}
\vspace{-0.5cm}
\caption{The time-domain waveforms for asymmetric Damour-Solodukhin wormholes composed of two delta-function effective potential barriers with unequal magnitudes.  
The waveforms at $x=x_0=\frac14$ are computed using the Green's functions for RSM and echo modes shown in Fig.~\ref{fig_doublesquare_asymmetric_GF}, for a source term given by Eq.~\eqref{pulseS}.  
The results for the RSMs appear in the left panel, while those for the echo modes are shown in the right panel.
The real and imaginary parts of the waveforms $\Psi$ are plotted versus time on a logarithmic scale, clearly revealing the ``echo'' feature, while the insets show the waveforms on the original linear scale.  
The echo period, visible in all four panels, corresponds to the spacing between successive asymptotic modes along the real frequency axis: $T=\frac{2\pi}{\Delta\omega}=4x_c$.  
The reflectionless scattering waves exhibit greater amplitude than those associated with the QNMs.  
As discussed in the text, this arises because the RSMs lie closer to the real frequency axis than the QNMs.}
\label{fig_doublesquare_waveforms_RSM_QNM}
\end{figure}

The resulting time-domain waveforms are presented in Fig.~\ref{fig_doublesquare_waveforms_RSM_QNM}.  
Unlike the Green's functions, the time-domain waveforms closely resemble one another.  
As expected, both waveforms exhibit echoes with period $T=4x_c$, corresponding to the round-trip time between the two delta peaks.  
This is also consistent with $T={2\pi}/{\Delta\omega}$, where ${\Delta\omega}$ is the spacing along the real frequency axis between successive asymptotic modes governed by Eqs.~\eqref{rootRSM} and~\eqref{exEstiQNM}, namely, $\Delta\omega={\pi}/{2x_c}$.  
As explained in Appx.~\ref{appE}, one can verify the correctness of these waveforms by extracting the underlying modes using the Prony method~\cite{agr-qnm-55}.  
For the echo waveforms, the dominant mode $\omega_0 = 1.54266 - 0.793539 i$ is found, in reasonable agreement with the value $1.54318 - 0.79374 i$ listed in Tab.~\ref{tabNumDblDelta}.  
For the RSM waveforms, one instead obtains $\omega_0 = 1.92975 - 0.261073 i$, consistent with the value $1.92285 - 0.260979 i$ given in Tab.~\ref{tabNumDblDelta}.  
In the present setup, owing to the toy-model nature of the potential, the echoes do not appear as an enveloping modulation with a shorter oscillation period primarily governed by the fundamental mode.  
This is because the oscillation period of the above extracted fundamental modes is much larger than the echo period, and thus is not visible in Fig.~\ref{fig_doublesquare_waveforms_RSM_QNM}.  
It is worth noting that the echoes do not correspond to any individual mode.  
This is also demonstrated by the Prony fit, since no mode is found with real part $\mathrm{Re}\,\omega \approx 2\pi/T \approx 6.28$.  
As emphasized in Refs.~\cite{agr-qnm-echoes-22, agr-qnm-echoes-20, agr-qnm-echoes-45, agr-qnm-continued-fraction-40}, the echoes arise as a collective effect due to a sequence of modes evenly distributed along the real frequency axis.  

When comparing the waveforms associated with the RSMs to those of the QNMs, one observes that the reflectionless scattering waves exhibit larger amplitudes than the QNM contributions.  
In the present case, the magnitude of the RSM waveforms exceeds that of the echo modes by about one order of magnitude.  
This behavior is understood to be general, since the RSM spectrum lies closer to the real frequency axis than that of the echo modes, as clearly demonstrated in Tab.~\ref{tabNumDblDelta}.  
As a result, despite their similarity, RSMs have a more significant impact on the time-domain waveforms.  

\section{Concluding remarks}\label{sec6}

In conclusion, this work explored the concept of RSMs in the context of asymmetric Damour-Solodukhin wormholes. 
By promoting from the real axis to the complex plane, we elaborate on the underlying similarity between the reflectionless and echo mode spectra.  
For asymptotic modes satisfying $|\mathrm{Re}\omega| \gg |\mathrm{Im}\omega|$, it is shown that these two spectra exhibit a strong resemblance, featuring an approximately uniform distribution parallel to the real frequency axis with the same spacing between successive modes. 
In particular, it was shown that the real parts of asymptotic QNMs coincide with those of RSMs.
While echo modes typically possess non-vanishing imaginary parts, the RSMs of Damour-Solodukhin wormholes mainly lie close to the real frequency axis, where the distance between the RSMs and the real frequency axis measures the degree of deviation from a perfectly symmetric Damour-Solodukhin.
Compared with echo modes, this property of RSMs leads to more substantial amplitudes in the time-domain waveforms, and consequently stronger echoes.  
This is demonstrated by numerically evaluating the waveforms associated with the two types of modes for a given identical source.  
We carry out the derivations using two complementary approaches, based on the scattering matrix and the Green's function.  

We argue that such an analysis for asymmetric wormholes is pertinent, since realistic wormholes, if they exist at all, are unlikely to be symmetric, and poles lying closer to the real frequency axis exert a more significant impact on time-domain waveforms.
Recent developments on gravitational wave echoes, spectral instability, echo modes, and Regge poles are closely related to, and partly motivated by, the ongoing efforts in black hole spectroscopy.
Given the arguments that greybody factors may provide more relevant observables in light of black hole spectral instability, the notion of RSMs was proposed and explored.
The present study further elaborates on this concept, primarily in the context of Damour-Solodukhin wormholes, and offers an explicit comparison with echo modes.
It is worth noting that recently in~\cite{agr-qnm-instability-85}, the RSMs, referred to by the authors as total absorption or virtual absorption modes, were investigated for the Regge–Wheeler effective potential, for which the algebraically special mode is a well-known solution, and also in higher dimensions.
It is understood that echo modes and RSMs are two perspectives that complement one another and together provide effective tools for describing the underlying phenomenon.

\section*{Acknowledgements}

We are thankful for insightful discussions with Michael D. Green, Jodin C. Morey, and Guan-Ru Li.
We gratefully acknowledge the financial support from Brazilian agencies 
Funda\c{c}\~ao de Amparo \`a Pesquisa do Estado de S\~ao Paulo (FAPESP), 
Funda\c{c}\~ao de Amparo \`a Pesquisa do Estado do Rio de Janeiro (FAPERJ), 
Conselho Nacional de Desenvolvimento Cient\'{\i}fico e Tecnol\'ogico (CNPq), 
and Coordena\c{c}\~ao de Aperfei\c{c}oamento de Pessoal de N\'ivel Superior (CAPES).
This work is supported by the National Natural Science Foundation of China (NSFC).
A part of this work was developed under the project Institutos Nacionais de Ci\^{e}ncias e Tecnologia - Física Nuclear e Aplica\c{c}\~{o}es (INCT/FNA) Proc. No. 464898/2014-5.
This research is also supported by the Center for Scientific Computing (NCC/GridUNESP) of São Paulo State University (UNESP).

\appendix

\section{Flux conservation and the nonvanishing imaginary parts of the reflectionless modes}\label{appA}

In this appendix, we provide a brief account of the sign of RSMs' imaginary parts and the flux conservation.
The derivations are essentially based on the properties of the master equation and the relevant boundary conditions, which closely follow those given by Andersson and Thylwe~\cite{agr-qnm-Regge-02} in their analysis of the black hole Regge poles.
One left-multiplies the master equation Eq.~\eqref{eq2} by the complex conjugate of the wavefunction $\Psi^\dagger=\Psi^\dagger(\omega, x)$, and then subtracts from it the complex conjugate of the resulting equation.
Given that the effective potential is a real function, we have
\bqn
\Psi^\dagger\frac{d^2\Psi}{dx^2}-\Psi\frac{d^2\Psi^\dagger}{dx^2}=-2i|\Psi|^2{\mathrm{Im}(\omega^2)} .\label{Wprime}
\eqn
Meanwhile, one can estimate the asymptotic values of the Wronskian using Eq.~\eqref{def_g} and the definition Eq.~\eqref{RefTransA}.
Specifically, we have
\begin{equation}
\Psi \sim
\begin{cases}
   \mathcal{T} e^{-i\omega x}, &  x \to -\infty, \\
   e^{-i\omega x} + \mathcal{R} e^{+i\omega x} . &  x \to +\infty .
\end{cases}
\label{master_bc_in}
\end{equation}
At the boundary $x=x_\pm\to \pm\infty$, one can evaluate the Wronskians between $\Psi^\dagger$ and $\Psi$ as
\bqn
W_{x\to -\infty}= -2i\mathrm{Re}\omega_n \left|\mathcal{T}\right|^2 e^{2\mathrm{Im}\omega x^-},\label{Wneg}
\eqn
and
\bqn
W_{x\to +\infty}= -2i\ \mathrm{Re}\omega\ e^{2\mathrm{Im}\omega x^+} 
+2i\ \mathrm{Re}\omega \left|\mathcal{R}\right|^2 e^{-2\mathrm{Im}\omega x^+}
+4\mathrm{Im}\omega\ \mathrm{Im}\left[\mathcal{T}^* e^{-2i\mathrm{Re}\omega x^+}\right].\label{Wpos}
\eqn

It is noted that the difference between Eqs.~\eqref{Wpos} and~\eqref{Wneg} furnishes the integration of Eq.~\eqref{Wprime}, and one finds,
\bqn
e^{-2\mathrm{Im}\omega x^+}\left|\mathcal{R}\right|^2+e^{2\mathrm{Im}\omega x^-}\left(\left|\mathcal{T}\right|^2-1\right)
=
2i\frac{\mathrm{Im}\omega}{\mathrm{Re}\omega}\ \mathrm{Im}\left[\mathcal{R}^* e^{-2i\mathrm{Re}\omega x^+}\right]
-2\int_{x^-}^{x^+} \left|\Psi\right|^2{\mathrm{Im}\omega}\ dx   . \label{modFluxCon}
\eqn
Taking the limits of the integration $r_*^-\to -\infty$ and $r_*^+\to \infty$, Eq~\eqref{modFluxCon} readily falls back to the flux-conservation $|\mathcal{R}|^2+|\mathcal{T}|^2=1$ as long as $\mathrm{Im}\omega=0$, which is the scenario discussed in~\cite{agr-qnm-instability-63}.
However, the condition no longer holds for the reflectionless modes with a complex frequency $\omega=\omega_n$.
Specifically, for $\mathcal{R}=0$, Eq.~\eqref{modFluxCon} simplifies to
\bqn
\left|\mathcal{T}\right|^2
=1-2e^{-2\mathrm{Im}\omega x^-}\int_{-\infty}^{+\infty} \left|\Psi\right|^2{\mathrm{Im}\omega}\ dx   , \label{reflecTT}
\eqn
which does not vanish once $\mathrm{Im}\omega_n\ne 0$.
For $\mathrm{Im}\omega_n \ne 0$, we have $\left|\mathcal{T}\right| \ne 1$.

\section{An alternative derivation of the reflection amplitude at the wormhole's throat}\label{appB}

Here, we provide a second derivation of Eqs.~\eqref{TildeRForm} and~\eqref{TildeCForm}, which is more intuitive but somewhat tedious.

To accommodate the wormhole metric, we recall the asymptotic forms of the wavefunctions for the black hole on the r.h.s. of the throat
\begin{equation}
f_1(\omega, x)=\left\{\begin{matrix}e^{-i \omega(x-x_c)}&x\to \ \mathrm{throat} \\A_{\mathrm{out}}e^{i\omega (x-x_c)}+A_{\mathrm{in}}e^{-i\omega (x-x_c)} &x\to +\infty\end{matrix}\right. ,\tag{\ref{Asf1}}
\end{equation}
and
\begin{equation}
g_1(\omega, x)=\left\{\begin{matrix}e^{i\omega (x-x_c)}&x\to +\infty \\B_{\mathrm{out}}e^{i \omega (x-x_c)}+B_{\mathrm{in}}e^{-i \omega (x-x_c)} &x\to  \ \mathrm{throat}\end{matrix}\right. ,\tag{\ref{Asg1}}
\end{equation}
where we have denoted the limit for the ingoing wave at $x\to \ \mathrm{throat}$ instead of $x\to -\infty$.

To assess the waveform on the other end of the throat, one performs a spatial reflection $x\to -x$ to the effective potential $V^\mathrm{L}_\mathrm{BH}$, then applies a shift to the left $x\to x+x_c$.
We have
\bqn
g_3(\omega, x)=\left\{\begin{matrix}e^{+i \omega(x+x_c)}&x\to  \ \mathrm{throat} \\{A'}_{\mathrm{out}}e^{-i\omega (x+x_c)}+{A'}_{\mathrm{in}}e^{i\omega (x+x_c)} &x\to -\infty\end{matrix}\right. ,\label{Asg3}
\eqn
and
\bqn
f_3(\omega, x)=\left\{\begin{matrix}e^{-i\omega (x+x_c)}&x\to -\infty \\{B'}_{\mathrm{out}}e^{-i \omega (x+x_c)}+{B'}_{\mathrm{in}}e^{i \omega (x+x_c)} &x\to  \ \mathrm{throat}\end{matrix}\right. ,\label{Asf3}
\eqn
where we have mapped $f_1\to g_3$ and $g_1\to f_3$, the prime indicates that it is a different black hole and, again, we denote the limit for the outgoing wave at $x\to  \ \mathrm{throat}$.

Now, on the one hand, we rewrite the ingoing wave at the wormhole's throat according to 
\begin{equation}
f_3(\omega,x) = \mathcal{D}\left[f_1(\omega,x)+{\mathcal{C}}(\omega)g_1(\omega,x)\right] .\tag{\ref{h1Cform}}
\end{equation}
By matching the ratios of the corresponding coefficients of the wavefunctions while making use of their asymptotic forms given in Eqs.~\eqref{Asf1},~\eqref{Asg1}, and~\eqref{Asf3}, we have
\bqn
\frac{e^{i \omega x_c}+{\mathcal C}  B_{\mathrm{in}}e^{i \omega x_c}}{{B'}_{\mathrm{out}} e^{-i \omega x_c}}=\frac{{\mathcal C} B_{\mathrm{out}} e^{-i \omega x_c}}{{B'}_{\mathrm{in}} e^{i \omega x_c}} ,
\eqn
which gives
\bqn
 {\mathcal C}
=\frac{{B'}_{\mathrm{in}}e^{2i\omega x_c}}{B_{\mathrm{out}}{B'}_{\mathrm{out}}e^{-2i \omega x_c} -B_{\mathrm{in}}{B'}_{\mathrm{in}}e^{2i\omega x_c}}
=\frac{{\mathcal{T}}^\mathrm{R}_\mathrm{BH}\bar{\mathcal{R}}^\mathrm{L}_\mathrm{BH}}{e^{-4i \omega x_c} - {\mathcal{R}}^\mathrm{R}_\mathrm{BH}\bar{\mathcal{R}}^\mathrm{L}_\mathrm{BH}} ,
\eqn
which is precisely Eq.~\eqref{TildeCForm}.

On the other hand, $f_3$ can be expressed in terms of the definition of the reflection amplitude given by Eq.~\eqref{h1Rform}
\bqn
f_3(\omega,x)\propto e^{-i\omega (x-x_c)} +{\mathcal{R}}(\omega)e^{i\omega (x-x_c)} .\nb
\eqn
By comparing it against Eq.~\eqref{Asf3}, one finds
\bqn
{\mathcal{R}}
= \frac{{B'}_{\mathrm{in}}}{{B'}_{\mathrm{out}}}e^{4i\omega x_c}
=\frac{\left(T^\mathrm{L}\right)^{-1}_{12}}{\left(T^\mathrm{L}\right)^{-1}_{11}}e^{4i\omega x_c},\nb
\eqn
which is Eq.~\eqref{TildeRForm} in the main text, derived using the transfer matrix.

\section{Pole cancellation in the Green's function framework}\label{appC}

In this appendix, we show that the numerator of the Green's function Eq.~\eqref{Gtilde_h3} does not become divergent at the black hole's quasinormal frequencies, $\omega\to \omega_n^\mathrm{R}$.

One can also evaluate the normalization $\mathcal{D}(\omega)$ by matching the coefficient of $e^{-i\omega x}$ at the throat.
Specifically, for QNMs, we have
\bqn
{B'}_{\mathrm{out}}e^{-i\omega x_c} = \mathcal{D}(\omega)\left[e^{i\omega x_c} +\mathcal{C}(\omega)B_{\mathrm{in}}e^{i\omega x_c}\right] ,
\eqn
thus
\bqn
\mathcal{D}(\omega)= \frac{{B'}_{\mathrm{out}}}{1 +\mathcal{C}(\omega)B_{\mathrm{in}}}e^{-2i\omega x_c} ,
\eqn
which diverges when the frequency attains those QNMs.
At such a limit, $\mathcal{C}(\omega)\to -1/B_\mathrm{in}$ remains finite, which is readily verified by substituting the condition Eq.~\eqref{rQNM}, $T_{21}^\mathrm{R}=B_\mathrm{out}=0$, into Eq.~\eqref{TildeCForm}.
Moreover, it is straightforward to show that the following ``$0\cdot\infty$'' type products have well-defined limits
\bqn
\lim\limits_{\omega\to\omega_n^\mathrm{R}}\left(\frac{1}{B_{\mathrm{in}}}+\mathcal{C}(\omega)\right)\mathcal{D}(\omega) = \frac{{B'}_{\mathrm{out}}}{B_{\mathrm{in}}}e^{-2i\omega x_c} ,\label{limCBin}
\eqn
and
\bqn
\lim\limits_{\omega\to\omega_n^\mathrm{R}}B_\mathrm{out}\mathcal{D}(\omega) = -B_{\mathrm{in}}{B'}_{\mathrm{in}}e^{2i\omega x_c}.\label{limDBout}
\eqn

Now, we proceed to evaluate the numerator of Eq.~\eqref{Gtilde_h3}, which reads
\bqn
&& f_3(x_<)g_1(x_>) \nb\\
&=&\left[\mathcal{D}(\omega)f_1(x_<)+\mathcal{C}(\omega)\mathcal{D}(\omega)g_1(x_<)\right]g_1(x_>) \nb\\
&\simeq&\left[\mathcal{D}(\omega)f_1(x_<)+\mathcal{C}(\omega)\mathcal{D}(\omega)\left(B_{\mathrm{in}}f_1(x_<)+B_{\mathrm{out}}g_3(x_<)e^{-2i\omega x_c}\right)  \right]g_1(x_>) \nb\\
&\simeq& \left\{\left[\mathcal{D}(\omega)f_1(x_<)+\mathcal{C}(\omega)\mathcal{D}(\omega)B_{\mathrm{in}}f_1(x_<)\right]+\mathcal{C}(\omega)\mathcal{D}(\omega)B_{\mathrm{out}}g_3(x_<)e^{-2i\omega x_c}\right\}g_1(x_>) ,\nb\\
&\to&  \left[{B'}_{\mathrm{out}}e^{-2i\omega x_c}f_1(x_<) + {B'}_{\mathrm{in}}g_3(x_<)\right]g_1(x_>)
\label{devDivCancel}
\eqn
where one estimates the wavefunction in the asymptotic throat region by utilizing the specific form Eq.~\eqref{Asg1}.
As the frequency takes the values of QNMs, one observes that the divergence in the first two terms in the bracket on the last line of Eq.~\eqref{devDivCancel} cancels out, due to the limit Eq.~\eqref{limCBin}.
The remainder term is also manifestly finite, as its value can be estimated by utilizing the limit~\eqref{limDBout}.

\section{Transfer matrix for double square potential barrier}\label{appD}

For the double square potential barrier Eq.~\eqref{VsquareCase}, the transfer matrix is obtained by the boundary conditions at $x=\pm W/2$.
For a given incident wave, there are four equations, derived from the continuity of the wavefunctions and their first-order derivatives, that govern the four unknown wave amplitudes.
Their ratios to that of the incident wave give
\bqn
T 
=\frac{1}{4\omega\sqrt{\omega^2-V^\mathrm{L,R}}}
\begin{pmatrix}\bar{T}_{11}&\bar{T}_{12}\\ \bar{T}_{21}&\bar{T}_{22}\end{pmatrix} , \nb\\ \label{Tsquare}
\eqn
where
\bqn
\bar{T}_{11}&=& -{e^{-iW^\mathrm{L,R}(\omega+\sqrt{\omega^2-V^\mathrm{L,R}})}\left[\left(\omega-\sqrt{\omega^2-V^\mathrm{L,R}}\right)^2+\left(\omega+\sqrt{\omega^2-V^\mathrm{L,R}}\right)^2 e^{2i\sqrt{\omega^2-V^\mathrm{L,R}}W^\mathrm{L,R}}\right]},\nb\\
\bar{T}_{12}&=& {\left(e^{iW^\mathrm{L,R}\sqrt{\omega^2-V^\mathrm{L,R}}}-e^{-iW^\mathrm{L,R}\sqrt{\omega^2-V^\mathrm{L,R}}}\right)V^\mathrm{L,R} },\nb\\
\bar{T}_{21}&=& -{\left(e^{iW^\mathrm{L,R}\sqrt{\omega^2-V^\mathrm{L,R}}}-e^{-iW^\mathrm{L,R}\sqrt{\omega^2-V^\mathrm{L,R}}}\right)V^\mathrm{L,R} } ,\nb\\
\bar{T}_{22}&=& {e^{iW^\mathrm{L,R}(\omega-\sqrt{\omega^2-V^\mathrm{L,R}})}\left[\left(\omega+\sqrt{\omega^2-V^\mathrm{L,R}}\right)^2+\left(\omega-\sqrt{\omega^2-V^\mathrm{L,R}}\right)^2 e^{2i\sqrt{\omega^2-V^\mathrm{L,R}}W^\mathrm{L,R}}\right]} .
\eqn
which further simplifies to Eq.~\eqref{TsquareSimplified}.

\section{Numerical integration for the waveform and the Prony method}\label{appE}

In this appendix, we derive explicit expressions of the Green's functions for asymmetric Damour-Solodukhin wormholes composed of two delta-function effective potential barriers and use them to evaluate the corresponding waveforms.
Moreover, we assess the accuracy of the numerical results by extracting the underlying modes using the Prony method.

Using the definition of the scattering matrix, substituting $V_0^\mathrm{L}=2$, $V_0^\mathrm{R}=1$, and $x_c=\frac12$ into Eqs.~\eqref{Tdelta} and~\eqref{TdeltaInv}, it is straightforward to find
\bqn
f(\omega, y) = \frac{1}{\omega}\left[i e^{i\left(\frac12+y\right)\omega}+(i+\omega)e^{-i\left(\frac12+y\right)\omega} \right], 
\eqn
\bqn
\widetilde{f}(\omega, y) = \frac{1}{\omega}\left[i (-i+\omega)e^{i\left(\frac12+y\right)\omega}-ie^{-i\left(\frac12+y\right)\omega} \right], 
\eqn
and
\bqn
g(\omega, x) = \frac{1}{2\omega}\left[(i+2\omega) e^{i\left(-\frac12+x\right)\omega}+ie^{-i\left(-\frac12+x\right)\omega}\right],
\eqn
for $-x_c < x,y < x_c$.

Subsequently, the Green's functions have the forms
\bqn
G(\omega, x, y) = \frac{1}{2\omega^2\left[(3i+2\omega)(i\cos\omega+\sin\omega )-2\mathrm{sinc}\omega\right]}e^{-i\omega(x+y)}\left[(i+\omega) + ie^{i\omega(1+2y) }\right]\left[i+(i+2\omega)e^{i\omega(-1+2x)} \right],\nb\\ \label{G2delta}
\eqn
and
\bqn
\widetilde{G}(\omega, x, y) = \frac{1}{2\omega^2(3\cos\omega+(2-i\omega)\mathrm{sinc}\omega)}e^{-i\omega(x+y)}\left[-i + (-i+\omega)e^{i\omega(1+2y) }\right]\left[i+(i+2\omega)e^{i\omega(-1+2x)} \right] ,\nb\\ \label{G2deltaTilde}
\eqn
for $-x_c < y < x < x_c$.

The time-domain waveforms are obtained by performing the inverse Fourier transformation:
\bqn
\Psi(t, x) = \int_{-\infty}^{\infty} d\omega e^{-i\omega x} \Psi(\omega, x) ,\label{invFourier}
\eqn
where $\Psi(\omega, x)$ is obtained by the integration over the source via Eq.~\eqref{omegaPsi}.

A few comments are in order regarding the numerator integration in Eq.~\eqref{invFourier}.  
First, due to the strong oscillations in the Green's functions Eqs.~\eqref{G2delta} and~\eqref{G2deltaTilde}, the convergence of the inverse Fourier transform is slow, even though $G\sim 1/\omega^2$ at large frequency.  
In practice, the numerical convergence can be improved by exploiting the freedom to choose a specific form of the source $S(\omega, y)$ to our advantage.  
Specifically, one considers a source term localized within a certain frequency window, such as the one proposed in Eq.~\eqref{pulseS}.  
This can be understood by noting that the integration along the real frequency axis can be recast as a sum of contributions from the residues in the lower half of the complex frequency plane, by employing Jordan's lemma.  
As the source is localized, it gives more weight to the residues closer to $\omega_0$, and therefore effectively facilitates convergence.  
The above polynomial form will apparently introduce an artificial residue at $\omega=\omega_0$, but one can always place it in the upper half of the complex plane so that it is not enclosed by the contour.  
For the same reason, although a more localized Gaussian perturbation $\exp\left(-(\omega-\omega_0)^2/2\sigma_0\right)$ may seem favorable, this form is not a valid choice because it invalidates Jordan's lemma.  
Also, the exponential factors in the numerator of the Green's functions may appear worrisome; however, they are ultimately suppressed by the factor $\exp(-i\omega t)$ as long as $t > |x \pm y|$, so that the contour integration can always be closed in the lower half-plane. 
From a physical viewpoint, this corresponds to the time domain when the initial perturbations reach the observer, so the observed signals are causally connected to the source~\cite{agr-qnm-29, agr-qnm-review-02}. 
In particular, once one factors out $\exp(-i \omega_n t)$ evaluated at the residues, the remaining terms are governed by these residues but are otherwise time-independent constants, and only the largest among them is relevant.  
This dominant contribution can be approximated by substituting the value of the residue (the fundamental mode) into the numerator. 
With the above setup, one finds that the integration over the entire real-frequency axis can be reasonably approximated by restricting to the range $\omega \in (-30, 30)$.

Subsequently, the Prony method~\cite{agr-qnm-55} is employed to extract the underlying modes of the time-domain waveform.  
The method decomposes a uniformly sampled signal into a sum of damped complex exponentials and provides estimates for the complex frequency and amplitude of the oscillation profiles. 
As discussed in the text, in our calculations we are able to extract the lowest-lying mode with satisfactory precision using this method.

\bibliographystyle{h-physrev}
\bibliography{references_qian}

\end{document}